\documentclass[final,5p,times,twocolumn]{elsarticle}

\usepackage{hyperref}
\hypersetup{breaklinks=true}

\usepackage{amssymb}
\usepackage{amsmath}
\usepackage{nth}
\usepackage{booktabs}
\usepackage{listings}

\usepackage{xcolor}
\definecolor{commentgreen}{RGB}{2,112,10}
\definecolor{eminence}{RGB}{108,48,130}
\definecolor{weborange}{RGB}{255,165,0}
\definecolor{frenchplum}{RGB}{129,20,83}

\lstnewenvironment{Python}
{\lstset {
language=Python,
frame=tb,
tabsize=4,
showstringspaces=false,
numbers=left,
breaklines=true,
commentstyle=\color{commentgreen},
keywordstyle=\color{eminence},
stringstyle=\color{blue},
basicstyle=\footnotesize\ttfamily, 
emph={int,char,double,float,unsigned,void,bool},
emphstyle={\color{red}},
escapechar=\&,
classoffset=1, 
otherkeywords={>,<,.,;,-,!,=,~},
morekeywords={>,<,.,;,-,!,=,~},
keywordstyle=\color{orange},
classoffset=0,
numberstyle=\scriptsize,numbersep=2.0pt
}}
{}

\lstnewenvironment{C}
{\lstset {
language=C,
frame=tb,
tabsize=4,
showstringspaces=false,
numbers=left,
breaklines=true,
commentstyle=\color{commentgreen},
keywordstyle=\color{eminence},
stringstyle=\color{red},
basicstyle=\footnotesize\ttfamily, 
emph={int,char,double,float,unsigned,void,bool},
emphstyle={\color{blue}},
escapechar=\&,
classoffset=1, 
otherkeywords={>,<,.,;,-,!,=,~},
morekeywords={>,<,.,;,-,!,=,~},
keywordstyle=\color{weborange},
classoffset=0,
numberstyle=\scriptsize,numbersep=2.0pt
}}
{}

\newcounter{bla}

\journal{Computer Physics Communications}

\begin{document}

\begin{frontmatter}



\title{OpenSBLI: Automated code-generation for heterogeneous computing architectures applied to compressible fluid dynamics on structured grids}


\author[a]{David J. Lusher\corref{author}}
\author[b]{Satya P. Jammy}
\author[a]{Neil D. Sandham}

\cortext[author] {Corresponding author.\\\texttt{E-mail address:} D.Lusher@soton.ac.uk}
\address[a]{Aerodynamics and Flight Mechanics group. University of Southampton. Boldrewood Campus, Southampton, SO16 7QF, United Kingdom}
\address[b]{Faculty of Mechanical Engineering. SRM University. AP, Andhra Pradesh, 522502, India}

\begin{abstract}
OpenSBLI is an open-source code-generation system for compressible fluid dynamics (CFD) on heterogeneous computing architectures. Written in Python, OpenSBLI is an explicit high-order finite-difference solver on structured curvilinear meshes. Shock-capturing is performed by a choice of high-order Weighted Essentially Non-Oscillatory (WENO) or Targeted Essentially Non-Oscillatory (TENO) schemes. OpenSBLI generates a complete CFD solver in the Oxford Parallel Structured (OPS) domain specific language. The OPS library is embedded in C code, enabling massively-parallel execution of the code on a variety of high-performance-computing architectures, including GPUs. The present paper presents a code base that has been completely rewritten from the earlier proof of concept (Jacobs et al, JoCS 18 (2017), 12-23), allowing shock capturing, coordinate transformations for complex geometries, and a wide range of boundary conditions, including solid walls with and without heat transfer. A suite of validation and verification cases are presented, plus demonstration of a large-scale Direct Numerical Simulation (DNS) of a transitional Shockwave Boundary Layer Interaction (SBLI). The code is shown to have good weak and strong scaling on multi-GPU clusters. We demonstrate that code-generation and domain specific languages are suitable for performing efficient large-scale simulations of complex fluid flows on emerging computing architectures.
\end{abstract}


\end{frontmatter}



{\bf PROGRAM SUMMARY}

\begin{small}
\noindent
{\em Program Title: OpenSBLI code-generation framework for compressible fluid dynamics on heterogeneous architectures}                                          \\
{\em Licensing provisions: GPLv3}                                   \\
{\em Programming languages: Python, C/C++, OPS DSL}                                   \\
{\em Nature of problem: The compressible 3D Navier-Stokes equations are solved via Implicit Large Eddy Simulation (ILES) or Direct Numerical Simulation (DNS).}\\
{\em Solution method: OpenSBLI [1,2] is a Python-based code-generation system that uses symbolic algebra to generate a complete CFD solver in C/C++. The basic algorithm is a stencil-based finite-difference solver on structured curvilinear meshes. Shock-capturing is performed by a selection of high-order Weighted/Targeted Essentially Non-Oscillatory (WENO/TENO) schemes. Explicit low-storage Runge-Kutta schemes are used for time-advancement.}\\
{\em Additional comments: The generated code is compliant with the Oxford Parallel Structured (OPS) [3] software library. OpenSBLI/OPS executables can be generated for the OpenMP, MPI, CUDA, OpenCL, and OpenACC parallel programming paradigms. Multi-GPU support is available via combinations of MPI with CUDA, OpenCL or OpenACC.}\\

\end{small}


\section{Introduction}\label{sec:introduction}
The role of Computational Fluid Dynamics (CFD) in modern aerospace research is well established \cite{spalart_venkatakrishnan_2016}. CFD has become an integral part of the aeronautical research and design process. CFD can complement the data obtained from wind tunnels and in-flight testing, at potentially a fraction of the cost. Furthermore, high-fidelity Large Eddy Simulations (LES) or Direct Numerical Simulations (DNS) can reveal physical insights that would be difficult to investigate experimentally. The dramatic increase in computational power over the past few decades has broadened the scope of problems that can be tackled by LES/DNS. Two of the main challenges in this field today, are the efficient utilization of computational hardware, and the development of accurate and reliable numerical methods.

A recent trend in high-performance computing (HPC), has been a shift to ever-increasing levels of heterogeneity \cite{top500}. Graphical Programming Units (GPUs) and other types of accelerators are now being applied to many diverse areas of computational science \cite{Nvidiacatalog}. In addition to the vast available compute capacities, these emerging architectures can offer substantial improvements in power efficiency for large systems. One of the drawbacks limiting their uptake compared to conventional CPU-based platforms however, is the need for programming models suited to these architectures. Existing CFD solvers designed for CPUs often contain large amounts of legacy code, and can be inflexible to changes in computational hardware. Porting existing codes to new architectures can be a very time consuming process. One potential solution to this problem is the \textit{`separation of concerns'} philosophy \cite{Ober2017}, applied in the present work. This approach separates the physical problem and numerical methods from their parallel implementation, allowing the researcher to focus solely on modelling the physical problem at hand, while benefiting from performance optimisations and hardware-specific knowledge from computer science \cite{Reguly2020}.

The present work describes the Python-based OpenSBLI automatic code-generation system for compressible fluid dynamics. OpenSBLI is a high-order stencil-based finite-difference solver on structured meshes. The symbolic algebra library SymPy \cite{sympy}, is used to generate a CFD solver tailored to the equations and schemes specified by the user in a high-level Python script. An initial proof-of-concept version of this approach was presented in \cite{JACOBS201712}, to demonstrate the basic feasibility of code-generation for CFD. The demonstrator version was limited to smooth (shock free) subsonic problems on triply periodic domains. Given the substantial changes needed to develop the concept into a more generally useful research code for shock-wave boundary-layer interactions, the new version of OpenSBLI in this work was started from a separate code base. It is capable of simulating complex wall-bounded flows with shockwaves, for supersonic and hypersonic CFD applications on curvilinear meshes. OpenSBLI generates a complete CFD solver in the Oxford Parallel Structured (OPS) Embedded Domain Specific Language (EDSL) \cite{OPSC,Reguly_2014_OPSC}. The OPS library enables execution of the code on multiple massively-parallel computing architectures. Performance of the automatically generated OPS code has been shown to as good as, or better than, hand-coded version of the same application \cite{OPS_perf,Mudalige2019}.


Recent examples of comparable CFD codes on GPUs includes: Hydra \cite{Reguly2016}, HiPSTAR \cite{Sandberg2015}, PyFR \cite{Witherden2014}, the HTR solver \cite{HTR_code}, and STREAmS \cite{bernardini2020streams}. Hydra is an unstructured mesh solver that has been widely used commercially for turbo-machinery applications. It uses the OP2 EDSL \cite{Mudalige2012OP2AA} from the same authors as the OPS library used in this work. HiPSTAR is a high-order curvilinear finite-difference code that also originated from the University of Southampton. It has been applied to large-scale simulations of both low and high-pressure turbine cascades. PyFR is a Python-based unstructured mesh framework, to solve advection-diffusion problems on streaming architectures. The code utilises a domain specific language that uses Mako templates for platform portability, allowing PyFR to be compatible with both the CUDA and OpenCL programming languages, and with OpenMP in C. HTR is a hypersonic aero-thermodynamics code that includes temperature-induced thermochemical effects, and a \nth{6} order Targeted Essentially Non-Oscillatory (TENO) scheme for shock-capturing. The code is written in the Regent programming language, using the task-based Legion system to execute the code on GPUs. Finally, STREAmS is a compressible DNS solver that uses a hybridised Weighted Essentially Non-Oscillatory (WENO) scheme for wall-bounded turbulent flows. Based on an existing CPU-solver, the code uses CUDA-Fortran90 kernels to run large-scale DNS on modern GPU clusters.

One of the novel features of OpenSBLI is the use of symbolic code-generation to write the simulation code from Python. The OPS library is then used to create parallel versions of the code for a number of parallel programming paradigms. At present this includes MPI, OpenMP, OpenMP+MPI, CUDA, OpenCL, and OpenACC. Code-generation allows for a large number of numerical schemes to be contained within a compact code-base. OpenSBLI has a number of high-order accurate spatial discretization schemes. These include various orders of WENO/TENO shock-capturing \cite{Shu1997,Borges2008,Fu2016,FU201797}, and central-differencing which can be cast in split skew-symmetric forms to improve numerical stability \cite{DUCROS2000114}. Time-advancement is performed by low-storage explicit \nth{3} and \nth{4} order Runge-Kutta schemes \cite{carpenter_kennedy_1994}. The ability to define equations compactly in index notation in a high-level Python script, gives the user the flexibility to control core aspects of the solver. A comparable project is that of Devito \cite{devito_main}, albeit for a different physical application. Devito also uses the SymPy Python library to generate finite-difference stencils. Aimed at geophysical applications, the symbolic library has been used to manipulate and optimise expressions to improve computational efficiency \citep{devito-compiler}. While initially targeting the Intel Xeon-Phi platform, Devito has recently utilised the OPS DSL used in this work to target GPU clusters \cite{devito_GPU}.

The purpose of this work is to describe the main features of the OpenSBLI design, with presentation of a suite of validation and verification test cases. The work is structured as follows: Section \ref{sec:numerical_methods} gives an overview of the numerical methods used in OpenSBLI. This includes a characteristic flux reconstruction, high-order WENO/TENO shock-capturing, central-differencing with one-sided boundary closures, and time-advancement schemes. Section \ref{sec:code_section} describes the OpenSBLI code implementation, with discussion of the core components and an example problem script. Section \ref{sec:OPS_section} shows examples of the OPS C code that results from the code-generation process. A selection of validation and verification cases are shown in section \ref{sec:results}, plus demonstration of a large-scale transitional Shockwave Boundary-Layer Interaction (SBLI) DNS in section \ref{sec:transitional_SBLI}. Finally, section \ref{sec:performance} gives a brief discussion of computational performance in the OPS DSL.

\section{Numerical methods}\label{sec:numerical_methods}
Scale-resolving simulations of turbulence benefit from the use of high-order accurate numerical methods, which can alleviate the excessive levels of numerical diffusion associated with lower-order approximations \cite{Ekaterinaris2005,Pirozzoli_review}. In the context of compressible turbulence with shockwaves, there are two contrasting requirements placed on the numerical methods. Shock-capturing schemes stabilise the solution by adding numerical dissipation in the vicinity of flow discontinuities, but have the detrimental effect of damping small-scale turbulence \cite{JOHNSEN20101213}. This creates a requirement for very fine grids to achieve acceptable resolution of small-scale flow structures, exacerbating the already high computational cost of LES/DNS.

Common approaches include the hybrid pairing of non-dissipative central schemes with shock-capturing schemes via a shock-sensor \cite{Pirozzoli_review}. While these have been successful, it is difficult to design shock-sensors suitable for every type of flow conditions \cite{JOHNSEN20101213}. WENO schemes \cite{Shu1997} are family of robust high-order shock-capturing methods that have seen widespread use in compressible CFD and Magneto-Hydrodynamics (MHD). Examples of their use for SBLI includes \cite{Gross2016}. WENO schemes have previously been shown to be superior to lower-order shock-capturing methods \cite{Tenaud2000,Brehm2015}. TENO schemes \cite{Fu2016,FU201797}, are a more recent development designed for reduced numerical dissipation compared to WENO \cite{Lusher2019_AIAA}. Alternative approaches includes the framework of \cite{Yee2018}, which pairs skew-symmetric central and Dispersion Relation Preserving (DRP) schemes to a dissipative WENO scheme applied as a non-linear filtering step. The non-dissipative part of the framework can be written in either the Ducros-split \cite{DUCROS2000114}, or entropy-split \cite{Yee2000_entropy} forms to enhance numerical stability. The present section describes the high-order numerical methods in OpenSBLI, beginning with the WENO/TENO flux reconstruction procedure for the compressible Navier-Stokes equations.

\subsection{Governing equations}\label{sec:governing_equations}
OpenSBLI uses numerical indices to distinguish between variables that have a dependence on dimension. For example, the Cartesian coordinate base $\left(x, y, z\right)$ and their respective velocity components $\left(u, v, w\right)$, are taken to be $\left(x_0, x_1, x_2\right)$, and $\left(u_0, u_1, u_2\right)$ respectively in the code. This approach allows the code-generation to be more flexible; all of the indexed quantities and loops are generated dynamically, based on the number of dimensions set in the problem script. The base governing equations in this work are the dimensionless compressible Navier-Stokes equations for a Newtonian fluid. Applying conservation of mass, momentum, and energy, in the three spatial directions $x_i$ $\left(i=0, 1, 2\right)$, results in a system of five partial differential equations to solve. These equations are defined for a density $\rho$, pressure $p$, temperature $T$, total energy $E$, and velocity components $u_k$ as
\begin{align}\label{ns_eqn}
\frac{\partial \rho}{\partial t} &+ \frac{\partial}{\partial x_k} \left(\rho u_k \right) = 0,\\
\frac{\partial}{\partial t}\left(\rho u_i\right) &+ \frac{\partial}{\partial x_k} \left(\rho u_i u_k + p \delta_{ik} - \tau_{ik}\right) = 0,\\
\frac{\partial}{\partial t}\left(\rho E\right) &+ \frac{\partial}{\partial x_k} \left(\rho u_k \left(E + \frac{p}{\rho}\right) + q_k - u_i \tau_{ik}\right) = 0,
\end{align}
with heat flux $q_k$ and stress tensor $\tau_{ij}$ defined as 
\begin{equation}\label{heat_flux}
q_k = \frac{-\mu}{\left(\gamma - 1\right) M_{\rm{ref}}^{2} Pr Re}\frac{\partial T}{\partial x_k},
\end{equation}
\begin{equation}\label{stress_tensor}
\tau_{ik} = \frac{\mu}{Re} \left(\frac{\partial u_i}{\partial x_k} + \frac{\partial u_k}{\partial x_i} - \frac{2}{3}\frac{\partial u_j}{\partial x_j} \delta_{ik}\right).
\end{equation}
$Pr$, $Re$, and $\gamma$ are the Prandtl number, Reynolds number and ratio of heat capacities respectively. The equations are non-dimensionalized by a reference velocity, density and temperature $\left(U^{*}_{\rm{ref}}, \rho^{*}_{\rm{ref}}, T^{*}_{\rm{ref}}\right)$. For the SBLI cases, the characteristic length is the displacement thickness $\delta^{*}$ of the boundary layer imposed at the inlet. Pressure is normalised by $\rho_{\rm{ref}}^{*} {U_{\rm{ref}}^{*}}^{2}$. For cases with temperature dependent dynamic viscosity, $\mu\left(T\right)$  is evaluated using Sutherland's law 
\begin{equation}\label{sutherland}
\mu\left(T\right) = T^{\frac{3}{2}} \left(\frac{1+\frac{T_{\textrm{s}}^{*}}{T_{\rm{ref}}^{*}} } {T + \frac{T_{\textrm{s}}^{*}}{T_{\rm{ref}}^{*}}}\right),
\end{equation}
for a reference temperature $T_{\rm{ref}}^{*}$. The Sutherland temperature constant is set to be $T_{\textrm{s}}^{*} = 110.4 \textrm{K}$. For a reference Mach number $M_{\rm{ref}}$, pressure and local speed of sound are defined as 

\begin{equation}\label{pressure_eqn}
p = \left(\gamma - 1\right) \left(\rho E - \frac{1}{2} \rho u_i u_i\right) = \frac{1}{\gamma M^{2}_{\rm{ref}}} \rho T \quad \textrm{and} \quad a = \sqrt{\frac{\gamma p}{\rho}}.
\end{equation}
For the SBLI cases, the skin friction $C_f$ is calculated from the wall shear stress $\tau_w$ as 
\begin{equation}\label{skin_friction_eqn}
\tau_w = \mu {\frac{\partial u}{\partial y}\vline}_{y=0}, \quad \quad C_f = \frac{\tau_w}{\frac{1}{2} \rho_{\rm{ref}} U^{2}_{\rm{ref}}}.
\end{equation}

The high-level Python interface in OpenSBLI allows users to modify the equations to be solved in the simulation. Examples of this are given in the code repository for the turbulent channel flow applications, where the governing equations are recast in split skew-symmetric formulations to improve numerical stability. It is expected that this flexibility will enable OpenSBLI to be extended to equations beyond the compressible Navier-Stokes equations presented in this work. The next section outlines the flux reconstruction applied to the convective terms in equations (1-3), for problems requiring the use of the WENO/TENO shock-capturing schemes. Heat-flux \eqref{heat_flux} and stress-tensor \eqref{stress_tensor} terms are computed with central differences, as described in section \ref{sec:central_schemes}. The system of equations is advanced in time using the explicit Runge-Kutta methods in section \ref{sec:rk_schemes}.

\subsection{Flux reconstruction}
The convective terms of the Navier-Stokes equations form a set of conservation laws that can be approximated by flux reconstruction methods. For a given physical dimension, a finite-difference method creates discrete representations of derivatives on a set of $i$ grid points as in figure \ref{fig:weno_stencil}. Taking the example of a scalar conservation equation

\begin{equation}\label{advection}
\frac{\partial U}{\partial t} + f(U)_x = 0, 
\end{equation}
the flux term $f(U)_x$ can be approximated by computing two half-node reconstructions $\left[\hat{f}_{i+ \frac{1}{2}}, \hat{f}_{i- \frac{1}{2}}\right]$, such that the flux $f(U)_x$ is replaced by 

\begin{equation}\label{FD_reconstruct}
\frac{1}{\Delta x} \left( \hat{f}_{i+\frac{1}{2}} - \hat{f}_{i-\frac{1}{2}}\right),
\end{equation}
for a grid spacing of $\Delta x$. For a general flux $f(U)$, splitting methods are applied to account for upstream and downstream propagating information
\begin{equation}\label{interface_flux}
f(U) = f^{+}(U) + f^{-}(U), 
\end{equation}
where the plus and minus superscripts represent the cases \cite{JIANG1996202}
\begin{equation}
\frac{\partial f^{+}(U)}{\partial U} > 0 \quad \textrm{and} \quad \frac{\partial f^{-}(U)}{\partial U} < 0.
\end{equation}
The most common flux-splitting method is the Local Lax-Friedrich (LLF) flux
\begin{equation}\label{eq:lax_scalar}
f^{\pm}(U) = \frac{1}{2} \left(f(U) \pm \alpha U\right),
\end{equation}
for a given wave-speed $\alpha$. For systems of equations, the flux can be applied to each component in succession. To improve the robustness of the shock-capturing \cite{Brehm2015}, the reconstructions are performed in characteristic space as described in \cite{Shu1997}. The algorithm is summarised as follows for a system of $j$ equations.

\begin{enumerate}
\item Construct the flux $f(U_j)$ and solution vector $U_j$ terms at every $i$ grid point.
\item At each half-node $x_{i+\frac{1}{2}}$ perform the following:
\begin{enumerate}
\item Compute the average $U_{i+\frac{1}{2}}$ state with either simple or Roe averaging.

\item Obtain the eigensystem $R(U_{i+\frac{1}{2}})$, $R^{-1}(U_{i+\frac{1}{2}})$ and $\Lambda(U_{i+\frac{1}{2}})$ to diagonalize the equations.
\item Transform the solution and flux vector into characteristic space as $V_j = R^{-1} U_j$, and $g_j = R^{-1}f(U_j)$. 
\item Apply the flux-splitting $g^{\pm}_{j} = \frac{1}{2} \left(g_j \pm \alpha_j V_j\right)$, where $\alpha_j = \textrm{max}_{k} \left\vert \lambda_j\right \vert$, are the characteristic wave-speeds over the local stencil points $k$.
\item Perform the high-order WENO/TENO reconstruction at $x_{i+\frac{1}{2}}$.
\item Transform the flux back to physical space $R \hat{g}^{\pm}_{i+\frac{1}{2}}$.
\end{enumerate}
\item Build the finite-difference approximation as in \eqref{FD_reconstruct}.
\item Repeat for all other dimensions in the problem.
\end{enumerate}
The half-node flux reconstructions $\left[\hat{f}_{i+ \frac{1}{2}}, \hat{f}_{i- \frac{1}{2}}\right]$ can take many forms, allowing for the construction of high-order approximations of the interface fluxes. OpenSBLI uses the WENO and TENO high-order reconstruction methods described in the next sections.

\subsection{Weighted Essentially Non-Oscillatory (WENO) schemes}
\begin{figure}
\begin{center}
\includegraphics[width=1.0\columnwidth]{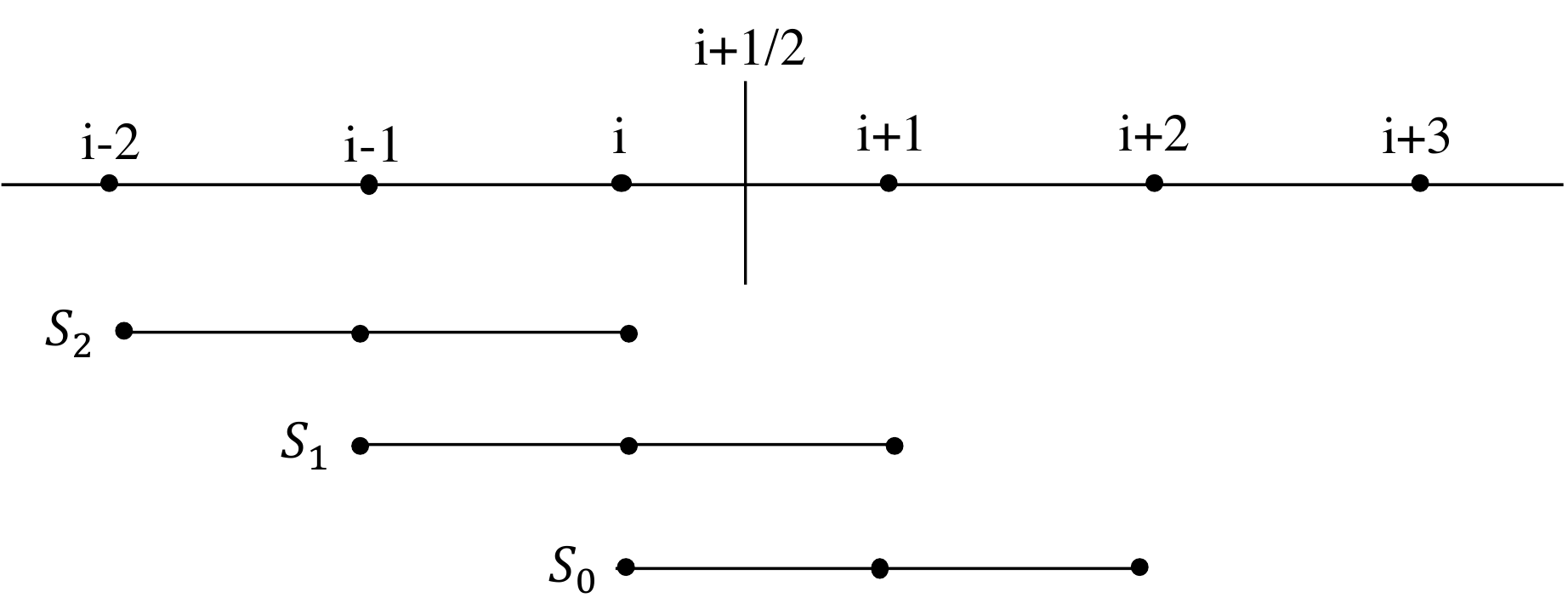}
\caption{Schematic of the finite-difference WENO stencils.}
\label{fig:weno_stencil}
\end{center}
\end{figure}

WENO schemes construct a high-order approximation for the half-node fluxes $\left[\hat{f}_{i+ \frac{1}{2}}, \hat{f}_{i- \frac{1}{2}}\right]$, by building up a convex combination of a set of smaller candidate stencils (figure \ref{fig:weno_stencil}). The candidate stencils are weighted based on the local smoothness of the flow. This mechanism avoids differencing over flow discontinuities, resulting in essentially non-oscillatory behaviour around shocks \cite{Shu1997}. For a WENO reconstruction of order $2k-1$, $k$ candidate stencils are required. In each candidate stencil an interpolation is applied such that 

\begin{equation}\label{eno_rhs}
\hat{f}^{(r)}_{i+ \frac{1}{2}} = \sum_{j=0}^{k-1} c_{rj} f_{i-r+j}, \quad \quad r = \left[0, k-1\right],
\end{equation}
where $c_{rj}$ are the standard ENO coefficients given in \cite{Shu1997}. The WENO reconstruction is then formed as
\begin{equation}\label{weno_rhs}
\hat{f}_{i+ \frac{1}{2}} = \sum_{r=0}^{k-1} \omega_r \hat{f}^{(r)}_{i+ \frac{1}{2}},
\end{equation}
for a non-linear weighting $\omega_r$. The choice of $\omega_r$ differs between the various WENO formulations. OpenSBLI uses both the original WENO-JS weightings \cite{Shu1997}, and those of the improved WENO-Z scheme \cite{Borges2008}. 

The fundamental aspect of ENO/WENO reconstructions is the mechanism to select certain candidate stencils. The scheme must be capable of identifying discontinuities in the flow, to remove discontinuity-crossing stencils from the final reconstruction. The most commonly used smoothness indicator was introduced by \cite{JIANG1996202}, defined as 

\begin{equation}\label{smoothness_indicator}
\beta_k = \sum_{l=1}^{r-1} \int_{x_{j-\frac{1}{2}}}^{x_{j+\frac{1}{2}}} \Delta x^{2l-1} \left(q_{k}^{(l)}\right)^{2} dx, 
\end{equation}
where $q_{k}^{(l)}$ is the $l$-th derivative of the $(r-1)$-th order interpolating polynomial $q_k (x)$ over a candidate stencil $S_k$. This smoothness indicator forms the sum of $L^2$ norms of the interpolating polynomial derivatives over a cell width. OpenSBLI generates the expressions for the smoothness indicators dynamically during code-generation, for the order specified by the user. The code supports generation of WENO schemes for any arbitrary odd order. As an example, for a \nth{5} order WENO scheme the smoothness indicators are given by
\begin{align}\label{fifth_smoothness}
\beta_0 &= \frac{13}{12} \left(f_i - 2f_{i+1} + f_{i+2}\right)^2 + \frac{1}{4}\left(3f_i - 4f_{i+1} + f_{i+2}\right)^2,\\
\beta_1 &= \frac{13}{12} \left(f_{i-1} - 2f_{i} + f_{i+1}\right)^2 + \frac{1}{4} \left(f_{i-1} - f_{i+1}\right)^2 ,\\
\beta_2 &= \frac{13}{12} \left(f_{i-2} -2f_{i-1} + f_i\right)^2 + \frac{1}{4}\left(f_{i-2} - 4f_{i-1} + 3f_i\right)^2,
\end{align}
where $f$ is one of the discrete flux terms from the governing equations.

To construct the $(2k-1)$-th order WENO approximation, the non-linear WENO weights are normalized for $r = [0, k-1]$ as 
\begin{equation}\label{omega}
\omega_r = \frac{\alpha_r}{\sum_{n=0}^{k-1} \alpha_n}, 
\end{equation}
with the smoothness indicators $\beta_r$ forming part of the alpha terms as 
\begin{equation}\label{alpha}
\alpha_r = \frac{d_r}{\left( \epsilon + \beta_{r} \right)^p}, 
\end{equation}
for optimal weights $d_r$ and constants, $p$, $\epsilon$. The standard value of $p=2$ is used, and $\epsilon$ set to a small non-zero number $(10^{-6})$ to avoid division by zero. Optimal weights $d_r$ are taken from \cite{Shu1997}.

\subsection{WENO-Z formulation}\label{wenoz_sect}
The WENO-Z formulation \cite{Borges2008}, \cite{Castro2011}, is a substantial improvement over the base scheme in terms of achieving lower numerical dissipation \cite{Lusher2019}, while retaining robust shock-capturing. Non-linear weights $\omega_r$ from the improved formulation are
\begin{equation}\label{wenoz_weights}
\omega^{z}_{r} = \frac{\alpha^{z}_{r}}{\sum^{k-1}_{n=0}\alpha^{z}_{n}}, \quad\quad \alpha_{r}^{z} = d_r \left(1 + \left(\frac{\tau}{\sigma_r + \epsilon}\right)^{2}\right),
\end{equation}
with ($\epsilon \sim 10^{-16}$). Smoothness indicators $\beta_r$ and optimal weights $d_r$ are unchanged. WENO-Z introduces a global smoothness measure $\tau$, representing the absolute difference in the smoothness indicators over the full reconstruction stencil. As an example, at \nth{5} order ($k=3$) the global smoothness measure is calculated from \cite{Castro2011} as $\tau = \lvert \beta_0 - \beta_2 \rvert$. As only minor changes are required compared to the base scheme, the performance impact of WENO-Z is almost negligible \cite{Lusher2019}.

\subsection{Targeted Essentially Non-Oscillatory (TENO) schemes}
\begin{figure}
\begin{center}
\includegraphics[width=1.0\columnwidth]{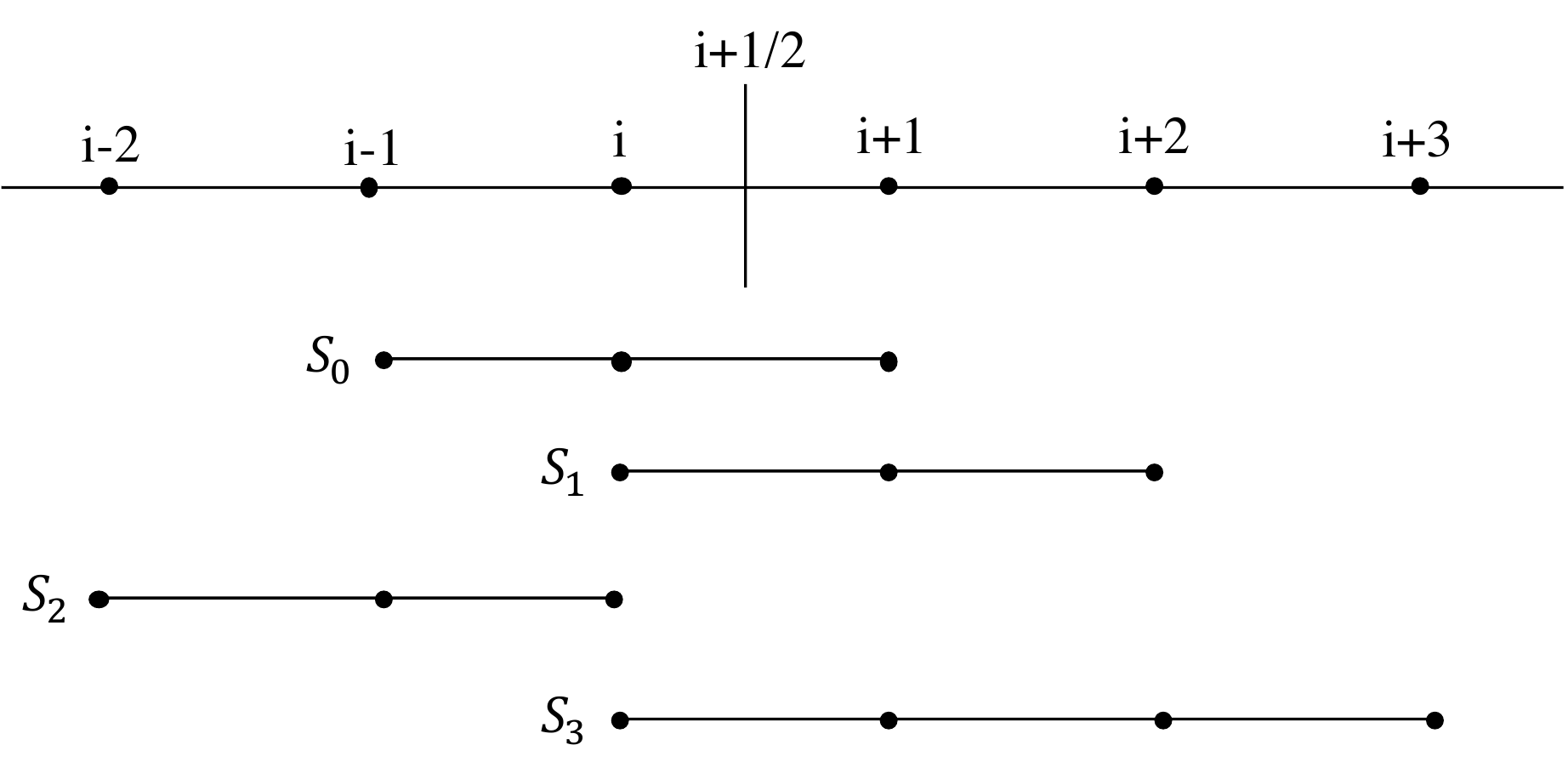}
\caption{Schematic of the finite-difference TENO stencils.}
\label{fig:teno_stencil}
\end{center}
\end{figure}
A more significant improvement was introduced by the TENO schemes of \cite{Fu2016,FU201797}. TENO schemes fit into the same flux reconstruction framework as WENO, with identical flux-splitting and characteristic decompositions. TENO schemes differ from WENO however in three fundamental ways: a staggered ordering of candidate stencils as in figure \ref{fig:teno_stencil}, the complete removal of candidate stencils deemed to be non-smooth, and modified non-linear weights optimized for low numerical dissipation. For a $K$-th order TENO scheme with $r$ candidate stencils, the non-linear weights take the form
\begin{equation}
\omega_r = \frac{d_r \delta_r}{\sum_{r=0}^{K-3} d_r \delta_r},
\end{equation}
where $\delta_r$ is a discrete cut-off function
\[
\delta_r = 
\begin{cases}\label{cutoff_eqn}
    0 & \text{if} \quad \chi_r < C_T \\
    1 & \text{otherwise}\\
\end{cases}
\]
for a tunable cut-off parameter $C_T$. The smoothness measures $\chi_r$ are the same as the weight normalization process as in WENO
\begin{equation}
\chi_r = \frac{\gamma_r}{\sum^{K-3}_{r=0}\gamma_r},
\end{equation}
comprised of the WENO-Z inspired form of non-linear TENO weights \citep{Fu2016}
\begin{equation}\label{TENO_weights}
\gamma_r = \left(C + \frac{\tau_K}{\beta_r + \epsilon}\right)^{q}, \quad r = 0,\dots,K-3,
\end{equation}
with $C=1$, and $q=6$. Smoothness indicators $\beta_r$ are unchanged from the standard Jiang-Shu formulation \citep{JIANG1996202}, and $\epsilon \sim 10^{-16}$. The global smoothness indicator $\tau_K$ measures smoothness over the entire stencil, and is given for \nth{5} and \nth{6} order TENO schemes as
\begin{align}\label{tau_5}
\tau_5 &= \lvert \beta_0 - \beta_2 \rvert,\\
\tau_6 &= \lvert \beta_3 - \frac{1}{6}\left(\beta_0 + \beta_2 + 4\beta_1\right) \rvert.
\end{align}
$C_T$ is the user-specified parameter which determines whether a given candidate stencil is rejected or contributes to the flux reconstruction. Lower values of $C_T$ are suitable for compressible turbulence simulations where minimal numerical dissipation is required, but this comes at the cost of increased spurious oscillations around shockwaves. $C_T$ is typically taken to be between $10^{-5}$ and $10^{-7}$, depending on the physical problem. The computational cost of the TENO schemes is approximately 15-20\% greater than a WENO scheme of equivalent order \cite{Lusher2019}, but offers significantly lower dissipation while retaining sharp shock capturing. OpenSBLI has both \nth{5} and \nth{6} order TENO schemes available.

\subsection{Central schemes for heat-flux, viscous, and metric terms}\label{sec:central_schemes}
Diffusive terms in OpenSBLI are computed by central-differences. Central schemes are also applied to smooth problems that do not require shock-capturing. The code-generation can produce central-difference approximations for any even-ordered central scheme. A fourth order central scheme is used throughout this work. For a grid spacing of $\Delta x$, the formula for first and second derivatives are

\begin{align}
f'_{i} &= \frac{-f_{i-2} + 8f_{i-1} - 8f_{i+1} + f_{i+2}}{12 \Delta x},\\
f''_{i} &= \frac{-f_{i-2} + 16f_{i-1} - 30f_{i} + 16f_{i+1} -f_{i+2}}{12 \Delta x^2}.
\end{align}

At non-periodic domain boundaries, the central-differences are replaced by one-sided derivatives. There are two \nth{4} order boundary schemes available in OpenSBLI, to maintain a consistent order throughout the domain. The first is the scheme of \cite{Carpenter1998}, which uses a [-5,5] stencil at domain boundaries. The second scheme \cite{Ekaterinaris2005} modifies two points at each boundary with a [-4,4] stencil such that

\begin{align}
f_{0}^{\prime} &=\frac{1}{12 \Delta x}\left(-25 f_{0}+48 f_{1}-36 f_{2}+16 f_{3}-3 f_{4}\right), \\ 
f_{1}^{\prime} &=\frac{1}{12 \Delta x}\left(-3 f_{0}-10 f_{1}+18 f_{2}-6 f_{3}+f_{4}\right), \\
f_{N-2}^{\prime} &= \frac{1}{12 \Delta x}\left(-f_{N-5}+6 f_{N-4}-18 f_{N-3} +10 f_{N-2}+3 f_{N-1}\right), \\
 f_{N-1}^{\prime} &= \frac{1}{12 \Delta x}\left(3 f_{N-5}-16 f_{N-4}+36 f_{N-3} -48 f_{N-2}+25 f_{N-1}\right).
\end{align}
For both boundary schemes the second derivatives are computed for the final two interior points from \cite{Carpenter1998} using the formula
\begin{align}
f^{\prime \prime}_{0} &= \frac{35f_0 - 104f_1 + 114f_2 - 56f_3 + 11f_4}{12 \Delta x^{2}},\\
f^{\prime \prime}_{1} &= \frac{11f_0 - 20f_1 + 6f_2 +4f_3 -f_4}{12 \Delta x^{2}}.
\end{align}

OpenSBLI also contains a metric transformation class to symbolically transform derivatives to a set of curvilinear coordinates $\xi(x,y,z)$, $\eta(x,y,z)$, and $\zeta(x,y,z)$. These are used for simulations containing stretched and curved meshes. The metric terms are evaluated by the same central interior and boundary schemes described in this section. Metric terms are computed once at the start of the simulation, to be multiplied into derivative terms to perform the coordinate transformation. Further discussion of the metric transformation in OpenSBLI is given in \cite{Lusher_thesis}.

\subsection{Explicit Runge-Kutta time-stepping}\label{sec:rk_schemes}
Large-scale DNS of the compressible Navier-Stokes equations has considerable memory requirements, making low-storage time-advancement schemes an attractive option to tackle challenging flows \citep{Kennedy2000}. Furthermore, explicit methods avoid having to compute the expensive inversion of systems required by implicit schemes. Explicit methods with structured meshes are well suited to modern computational hardware options such as GPUs, as they avoid performance issues related to poor data locality. OpenSBLI currently has two low-storage time-stepping schemes available: a standard \nth{3} order Runge-Kutta scheme in the form proposed by \cite{wray1990minimal}, and one following the work of \cite{WILLIAMSON198048}. The second formulation is used throughout this work, and has been implemented in OpenSBLI for \nth{3} and \nth{4} order, plus a \nth{3} order strong-stability-preserving (SSP) version to improve stability for flows containing discontinuities. Low-storage Runge-Kutta schemes require only two additional storage arrays per equation. For an $m$-stage scheme, time advancement of the solution vector $U$ from time level $U^n$ to $U^{n+1}$ is performed at stage $i=1, \dots, m$ such that
\begin{align}
dU^{(i)} &= A_i dU^{(i-1)} + \Delta t R\left(U^{(i-1)}\right),\\
U^{(i)} &= U^{(i-1)} + B_i dU^{(i)},\\
U^{n+1} &= U^{(m)},
\end{align}
for a constant time-step $\Delta t$, initial conditions $U^{(0)} = U^n$ and $dU^{(0)} = 0$, and residual $R(U^{(i-1)})$. The \nth{3} and \nth{4} order schemes have three and five sub-stages per iteration respectively. The coefficients $A_i$ and $B_i$ are taken from \cite{carpenter_kennedy_1994} for \nth{3} and \nth{4} order, and \cite{Gottlieb1998} for the SSP version of the \nth{3} order scheme.

\section{OpenSBLI code implementation}\label{sec:code_section}
\begin{figure}[!h]
\begin{center}
\includegraphics[width=1.0\columnwidth]{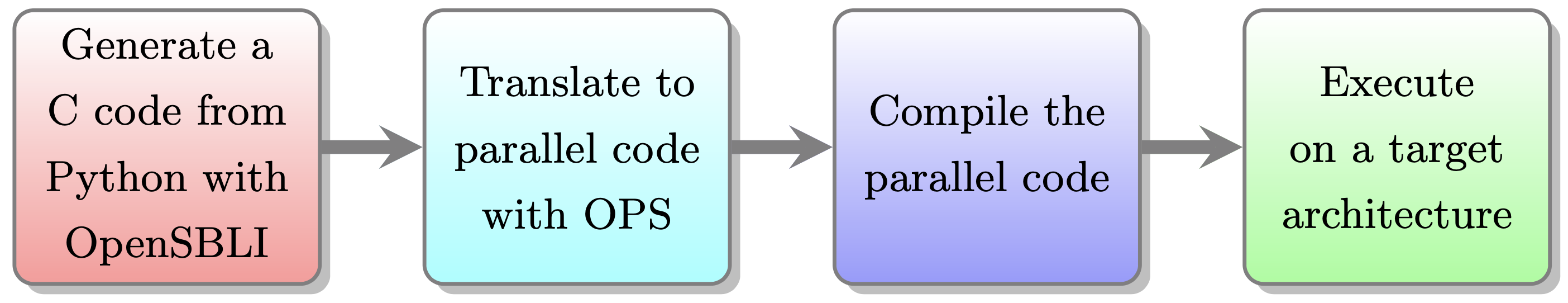}
\caption{The code-generation process in OpenSBLI/OPS.}
\label{fig:opensbli_flow_chart}
\end{center}
\end{figure}

This section outlines the main design principles of the OpenSBLI code-generation system, with examples from the high-level Python user interface. The purpose of OpenSBLI is to generate a complete, customizable, CFD solver in the OPS domain specific language \cite{OPSC,Reguly_2014_OPSC,OPS_perf,Mudalige2019}. OPS is a programming abstraction for massively-parallel computation on structured multi-block meshes. The OpenSBLI/OPS workflow is summarised in figure \ref{fig:opensbli_flow_chart}. In step 1, OpenSBLI is used to generate an OPS C code for a given physical problem. The OPS translator is then used to perform a source-to-source translation of the base code to a number of parallel programming paradigms. The parallel code can then be compiled and executed on a target architecture. This allows the code to be executed on a range of different hardware from a single source code.

It is important to note that a code-generation system is not a prerequisite to write OPS C code. As we will see in section \ref{sec:OPS_section} however, the standardised parallel templates of OPS code lend themselves well to a code-generation approach. Code-generation also enables far greater control of the resulting simulation code, leading to performance optimisations \cite{Jammy_Algo} that would be difficult in hand-written code. In addition, code-generation can improve code maintainability, by utilising an object-oriented design to share common functionality between classes. This approach allows for a large number of scheme options to be implemented in a relatively compact code base. 

The simulation code is tailored to the options selected by the user in the Python interface. Python was selected because it is an extremely versatile general-purpose programming language, with a rich ecosystem of external scientific libraries. One such example is the symbolic algebra library SymPy \cite{sympy}, which provides the fundamental building blocks of OpenSBLI. OpenSBLI has a number of symbolic data structures that inherit the functionality provided by SymPy. This functionality enables us to define, manipulate, and simplify symbolic expressions, using symbolic objects that follow the fundamental rules of mathematics. 

OpenSBLI parses and expands user-defined equations, before performing a symbolic discretization procedure for the selected numerical schemes. There are classes for boundary conditions, initial conditions, and handling of simulation input/output, which are all controlled by the user in the high-level problem script. Much of the complexity of code-generation comes from the need to make the abstraction as general as possible, so that it is flexible enough to target a wide range of applications within the domain of fluid dynamics. As the system has no inherent knowledge of the algorithms required by CFD, we have to apply sorting procedures to ensure dependencies are satisfied in the correct order in the resulting simulation code. The next section discusses the core components that make up the system.

\subsection{Creating an OpenSBLI problem script}\label{sec:opensbli_core}
\begin{figure}[!h]
\begin{center}
\includegraphics[width=1.0\columnwidth]{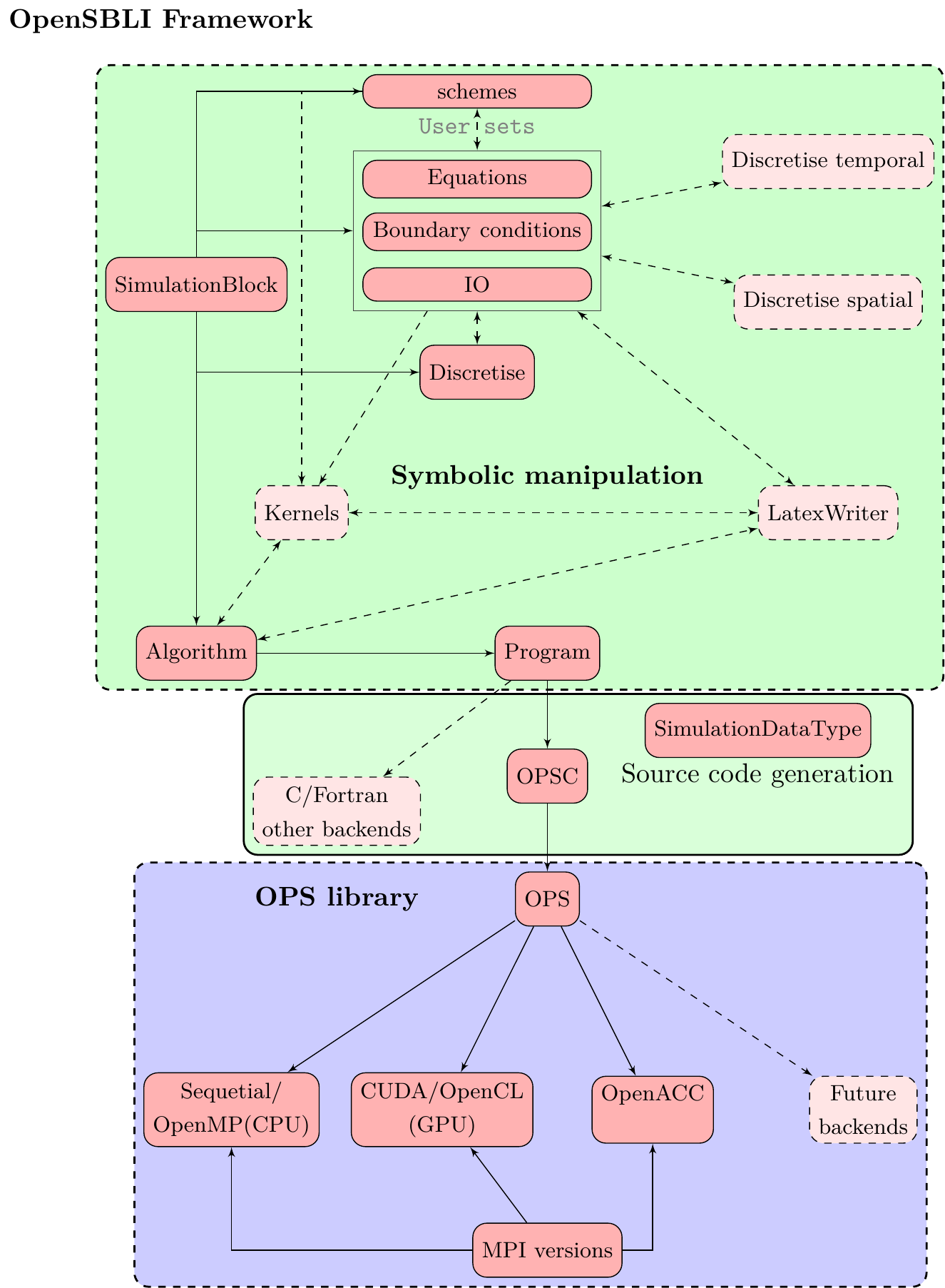}
\caption{Schematic of the major components in the OpenSBLI automatic source code generation framework.}
\label{fig:opensbli_design}
\end{center}
\end{figure}

OpenSBLI is comprised of a number of Python classes that provide the core functionality needed to generate a CFD solver. Figure \ref{fig:opensbli_design} illustrates these core components. A complete description of all of the source code files is contained in the user manual that accompanies this work. In the following sections, the core components of the system are described in the order that they would appear in a user-defined problem script.

\subsubsection{Defining and expanding equations}\label{sec:defining_equations}
\begin{Python}[caption={Defining the compressible Navier-Stokes equations in OpenSBLI.},captionpos=b]
ndim, scheme = 3, "**{\'scheme\':\'Teno\'}"
# Define the compresible Navier-Stokes equations in Einstein notation.
mass = "Eq(Der(rho,t), - Conservative(rhou_j,x_j,%s))" % scheme
momentum = "Eq(Der(rhou_i,t) , -Conservative(rhou_i*u_j + KD(_i,_j)*p,x_j , %s) + Der(tau_i_j,x_j) )" % scheme
energy = "Eq(Der(rhoE,t), - Conservative((p+rhoE)*u_j,x_j, %s) - Der(q_j,x_j) + Der(u_i*tau_i_j ,x_j) )" % scheme
stress_tensor = "Eq(tau_i_j, (mu/Re)*(Der(u_i,x_j)+ Der(u_j,x_i) - (2/3)* KD(_i,_j)* Der(u_k,x_k)))"
heat_flux = "Eq(q_j, (-mu/((gama-1)*Minf*Minf*Pr*Re))*Der(T,x_j))"
substitutions = [stress_tensor, heat_flux]
constants = ["Re", "Pr", "gama", "Minf"]
\end{Python}

The first step when creating a simulation in OpenSBLI is to specify the equations to be solved. The code listing shows how this is done for the compressible Navier-Stokes equations defined in section \ref{sec:governing_equations}. Taking the example shown in line 3, the SymPy equation class \texttt{Eq} is used with the OpenSBLI derivative-handling classes \texttt{Der} and \texttt{Conservative}, to define the continuity equation
\begin{equation}
\frac{\partial \rho}{\partial t} + \frac{\partial}{\partial x_j} \left(\rho u_j \right) = 0.
\end{equation}
Note that there is an optional argument `scheme' passed to the spatial derivative, to specify that this derivative should be computed with the TENO scheme. Alternatively the user could have specified the use of a `WENO' or `Central' scheme here. If no scheme is passed, the derivative class defaults to a central derivative. Additionally, there is a \texttt{Skew} class available for skew-symmetric formulations of the central schemes. Lines 4-7 defines the momentum, energy, stress tensor, and heat flux equations in the same manner. In the first line the user selected the number of dimensions to be 3, which will be used when expanding these equations over the repeated index $j$. The symbol $t$ is reserved to denote temporal derivatives. In line 8, the stress tensor and heat flux equations are stored in a list to be substituted into the momentum and energy equations. This optional substitution step is used to simplify the form of the equations. Note that if a term has more than one index, these should be separated by an underscore (e.g. $\tau_{ij}$ is written as \texttt{tau\_i\_j}). Finally, a list of simulation constants are defined in line 9, including Reynolds number, Prandtl number and the ratio of specific heat capacities.

\begin{Python}
# Expand the continuity equation
EE = EinsteinEquation()
continuity = EE.expand(mass, ndim, "x", substitutions, constants)
\end{Python}

The next step is to instantiate the \texttt{EinsteinEquation} class, which contains the functionality for parsing and expanding of equations. The class contains an \texttt{expand} method, which expands the equation over the number of dimensions of the problem. A coordinate base symbol of $x$ is given, along with any substitutions and constants as necessary. This parsing and expansion step is repeated for each of the governing equations. 
\begin{Python}
# Define and expand the pressure and viscosity relations
pressure = "Eq(p, (gama-1)*(rhoE - rho*(1/2)*(KD(_i,_j)*u_i*u_j)))"
viscosity = "Eq(mu, T**(1.5)*(1+SuthT/RefT)/(T+SuthT/RefT))"
P = EE.expand(pressure, ndim, coordinate_symbol, substitutions, constants)
mu = EE.expand(viscosity, ndim, coordinate_symbol, substitutions, constants)
\end{Python}

Next we have to define any relations that are required by the base equations, such as the dynamic viscosity relation \eqref{sutherland}, or the equation of state \eqref{pressure_eqn}. In this example, Sutherland's law \eqref{sutherland} and the pressure evaluation \eqref{pressure_eqn} are defined in the same manner as before, before being parsed and expanded by the \texttt{EinsteinEquation} class. A Kronecker Delta class \texttt{KD} is applied here for the pressure calculation, to sum over the square of the velocity components. OpenSBLI contains other objects providing index functionality, such as the Levi-Civita symbol.

\begin{Python}
# Store the expanded equations and relations
SE = SimulationEquations()
CR = ConstituentRelations()
SE.add_equations(continuity), CR.add_equations(P)
\end{Python}

The governing equations and their relations have now been defined, parsed, and expanded. To distinguish between time-advanced equations and the supporting relations, we must now group them as either \texttt{SimulationEquations}, or \texttt{ConstituentRelations}. To do this, the two OpenSBLI classes are instantiated, and each of the previously expanded equations are added by the \texttt{add\_equation} method. Here we are only adding one equation to each for brevity, but the method accepts lists of multiple equations at once.

\subsubsection{Selection of numerical schemes}
\begin{Python}[caption={Spatial and temporal scheme selection in OpenSBLI.},captionpos=b]
# Set the numerical schemes
schemes = {}
LLF = LLFTeno(order=6, averaging = RoeAverage([0, 1]))
cent = Central(4)
rk = RungeKuttaLS(3, formulation='SSP')
schemes[LLF.name], schemes[cent.name], schemes[rk.name] = LLF, cent, rk
\end{Python}

Having defined all of the base equations, the user must now select the numerical schemes they wish to use. A standard Python dictionary is created in line 2 to hold the schemes. Lines 3-5 instantiate the \nth{6} order LLF TENO scheme for shock-capturing with Roe-averaging, the \nth{4} order Central scheme for viscous/heat-flux terms, and a \nth{3} order SSP Runge-Kutta scheme for time-advancement. The final step in line 6 is to store the schemes in the Python dictionary in the key-value syntax, where the name of the scheme is used as the key. At this point no symbolic discretization of the equations has been performed, the schemes have only been initialized.

\subsubsection{Setting boundary and initial conditions}\label{sec:opensbli_BCS}
\begin{Python}
# Setting an initial condition
initial = GridBasedInitialisation()
initial.add_equations(initial_equations)
# Selecting a boundary condition
boundaries = [[0, 0] for t in range(ndim)]
direction, side = 1, 0
boundaries[direction][side] = IsothermalWallBC(direction, side, wall_condition)
\end{Python}
To generate a complete CFD solver, we must also specify the initial and boundary conditions for the problem. The simplified code extract shows how this is done. Line 2 instantiates the \texttt{GridBasedInitialisation} class, which will be executed once at the start of the simulation ($t=0$). The \texttt{initial\_equations} variable would be a list of conditions written in the same manner as the equations in section \ref{sec:defining_equations}. The \texttt{add\_equations} method is then used in line 3 to add this set of initial conditions to the class.

Line 5 creates a list of lists to store the boundary condition classes for the problem. For a problem of dimension \texttt{ndim}, there are \texttt{ndim*2} boundary conditions that must be set. These are distinguished by having a side of 0 or 1. In the example, an isothermal wall boundary condition has been selected for the \nth{0} (bottom) side in the \nth{1} ($y$) direction of the problem. Boundary conditions are enforced on conservative variables by using the ghost (halo) points required by the shock-capturing schemes. In the case of the isothermal wall boundary shown here, the no-slip condition is first enforced on the boundary plane as
\begin{equation}
\rho u=\rho v=\rho w=0.
\end{equation}
The \texttt{wall\_condition} variable is an expression for the total energy $\rho E$, set depending on the thermal properties of the wall in question. For a constant temperature $T_w$ isothermal wall the energy is set for a wall density $\rho_{\textrm{w}}$ as 
\begin{equation}\label{eq:wall_energy}
\rho E = \frac{\rho_{\textrm{w}}T_{\textrm{w}}}{M_{\rm{ref}}^{2} \gamma \left(\gamma -1\right)},
\end{equation}
where the wall density is obtained from solving the continuity equation \eqref{ns_eqn}. The ghost flow in the halo points $\left(i < 0\right)$ is enforced by extrapolating temperature from the interior and wall as
\begin{equation}
T_{-i} = (i+1)T_{\textrm{w}} - iT_1.
\end{equation}
Density in the halos is evaluated using the extrapolated temperature values and the wall pressure $p_{\textrm{w}}$ as 
\begin{equation}\label{eq:halo_density}
\rho_{-1} = \frac{M_{\rm{ref}}^{2} \gamma p_{\textrm{w}}}{T_{-i}}.
\end{equation}
Momentum components in the halos are set by reflecting the velocity components from the interior flow with a reversed sign such that
\begin{align}
\rho u_{-i} &= -\rho_{-i} u_{i},\\
\rho v_{-i} &= -\rho_{-i} v_{i},\\
\rho w_{-i} &= -\rho_{-i} w_{i},
\end{align}
where $\rho_{-i}$ is the halo density calculated in equation \eqref{eq:halo_density}. In the case of an adiabatic wall, the formula for a one-sided \nth{4} order approximation of $\partial T / \partial y = 0$ is rearranged using the interior points to calculate the unknown wall temperature $T_{\textrm{w}}$, to enforce the zero heat-flux condition. This wall temperature is then used to set the wall energy as in equation \eqref{eq:wall_energy}. In both cases the value of the wall density is left to float by letting the scheme solve the continuity equation, which avoids over-specifying the boundary condition. A full listing of the available boundary conditions is given in the user manual.

\subsubsection{Selecting HDF5 file I/O options}
\begin{Python}[caption={HDF5 input/output file writing in OpenSBLI.},captionpos=b]
kwargs = {'iotype': "Write"}
h5 = iohdf5(save_every=10000, **kwargs)
h5.add_arrays(SE.time_advance_arrays)
h5.add_arrays([DataObject('T')])
\end{Python}

All input/output of simulation data in OpenSBLI is handled by the parallel Hierarchical Data Format (HDF5) library \cite{hdf5}. HDF5 enables large data files to be stored and organised into groups containing multiple named datasets. In lines 1-2 of the example code, the OpenSBLI \texttt{iohdf5} class is instantiated with the write (output) argument. This class will control all of the output writing to disk. An optional argument \texttt{save\_every} is specified, to write intermediate simulation data, for example every 10,000 iterations. The \texttt{iohdf5} class has an \texttt{add\_arrays} method, which can be used to set which flow variables to write to disk. In this example the time-advance arrays $\left(\rho, \rho u, \rho v, \rho w, \rho E\right)$, and temperature $T$, have been selected for the output. Restarting of simulations or the reading of a coordinate mesh is also handled by this class, with the \texttt{iotype} `Read'.

\subsubsection{Creating a block and generating the C code}
\begin{Python}[caption={Creating an OPS C code in OpenSBLI.},captionpos=b]
# Create a simulation block
block = SimulationBlock(ndim, block_number = 0)
# Set the user options on the block
block.set_discretisation_schemes(schemes)
block.set_block_boundaries(boundaries)
block.setio(h5)
block.set_equations([CR, SE, initial])
# Begin the symbolic discretisation process
block.discretise()
# Create an algorithm to order the computations
alg = TraditionalAlgorithmRK(block)
SimulationDataType.set_datatype(Double)
# Generate the OPS C code and write it to file
OPSC(alg)
\end{Python}

Up until this point in the script, no symbolic discretization has been performed. The user has simply been selecting the options they require for the simulation. The core component that links everything together is called the \texttt{SimulationBlock}. A \texttt{SimulationBlock} is created in line 2, with the number of dimensions of the problem. We then set the numerical schemes, boundary conditions, HDF5 I/O, and all of the equations on the block. To begin the symbolic discretisation, we call the \texttt{block.discretise()} method. This method loops over all of the equations to be solved, and calls discretisation routines in each of the numerical methods where applicable. The continuous derivatives in the equations are converted into discrete representations, constructed from the symbolic objects that will be discussed in section \ref{sec:opensbli_objects}. These discrete equations are stored in computational \texttt{Kernels}, that are discussed in section \ref{sec:opensbli_kernels}.

Line 11 initialises an algorithm class based on the block. The algorithm class sets out the order that individual components should appear in the simulation code. For example, precedence is given to components involved in initialising the simulation, such as the declaration of global constants and memory allocation for storage arrays. Next would be the initial conditions to set from the \texttt{GridBasedInitialisation} class, before declaration of the main iteration and Runge-Kutta sub-stage loops and their components. At the end of the algorithm would be the calls to the HDF5 library to write the simulation output to disk. 

The final stages in lines 12 and 14 are to set the numerical precision of the simulation and begin the code-writing process with the \texttt{OPSC} class. The \texttt{OPSC} class is derived from the \texttt{C99CodeWriter} contained in SymPy \cite{sympy}. It contains methods that return C-compliant expressions for the discrete symbolic equations stored in the computational \texttt{Kernels}. Additionally, the \texttt{OPSC} class contains templates of calls to the OPS library functions, which are populated based on the computations stored in the \texttt{Kernels}. Examples of these OPS library functions in C are given in section \ref{sec:OPS_section}. Users would also input numerical values for simulation constants at this stage. Once the \texttt{OPSC} procedure is complete, a set of C/C++ codes are written out for translation and compilation with OPS as in figure \ref{fig:opensbli_flow_chart}. In practice, execution of the entire code-generation process takes less than a minute for the most demanding cases.

\subsection{Key OpenSBLI features and data structures}
In this section we briefly discuss the OpenSBLI data structures that are used internally during the code-generation process. A more complete guide is given in the accompanying user manual and in \cite{Lusher_thesis}. The final part of this section highlights some of the optimisations that are possible in the symbolic engine.

\subsubsection{Data structures}\label{sec:opensbli_objects}
Equations in OpenSBLI are constructed using the \texttt{Eq} class and the data structures outlined in this section. Equations can be defined either as strings to parse (as shown in section \ref{sec:defining_equations}), or by manually calling the following classes.
\begin{enumerate}
\item{\texttt{DataObject:} Variables to be stored as global arrays on the entire grid. These objects are not indexed, they represent terms in the continuous symbolic equations.}
\item{\texttt{DataSet (e.g. T[0,0,0]):} \texttt{SimulatonBlock}-specific versions of \texttt{DataObjects}. These objects have `ndim' indices, which are incremented per direction to create discrete finite-difference representations.}
\item{\texttt{GridVariable:} Temporary local variables that are used for intermediate calculations inside computational kernels. They contain a single value at a given grid point, to perform intermediate calculations.}
\item{\texttt{CoordinateObject:} Direction based coordinate objects. Used for defining derivatives and metric handling of the continuous equations.}
\item{\texttt{ConstantObject:} Used to define constant parameters. Examples would include Reynolds number, Prandtl number, and reference quantities.}
\end{enumerate}

\subsubsection{OpenSBLI Kernels}\label{sec:opensbli_kernels}
\begin{Python}
# Create an OpenSBLI Kernel
kernel = Kernel(block, computation_name="%s boundary direction%d side%d" % (bc_name, direction, side))
kernel = self.set_kernel_range(kernel, block)
kernel.add_equation(BC_equations)
\end{Python}

A key concept in OpenSBLI is that of the \texttt{Kernel}. The \texttt{Kernel} class is a way of holding the information required to perform a certain computation over a specified grid range. They are converted during the code-writing process into the OPS kernels discussed in section \ref{sec:OPS_section}. Kernels are created internally as shown in the example code, where a \texttt{Kernel} object is created in line 2 for a named boundary condition. In line 3 the spatial grid range for this computation is set using grid information stored in the \texttt{SimulationBlock}. The equations to evaluate over this grid range can then be added with the \texttt{add\_equation} method.

Each instance of the \texttt{Kernel} class extracts and stores information from its equations. This includes determining which of the terms in the equations require read, write, or read-write data access patterns. It must also inspect the relative data access of each quantity, to build up a set of numerical finite-difference stencils. For example, a \texttt{Kernel} for a \nth{4} order central difference would detect that it requires read access at $\left\{-2, -1, 1, 2\right\}$ in a certain direction. The \texttt{Kernel} would also contain a work array with write access at $\left\{0\right\}$, to store the value of the derivative at each grid location. The \texttt{Kernel} performs all of the `bookkeeping' work, that enables us to automatically generate the parallel OPS loops and C functions described in section \ref{sec:OPS_section}. All of this information would otherwise have to be maintained and updated manually for each computation in a static hand-written code, which can be time consuming and error prone.

\subsection{Oxford Parallel Structured (OPS) library}\label{sec:OPS_section}
As previously mentioned, automated parallel abstractions can offer researchers an easy way to utilise modern hardware. In addition to x86-CPU and GPU-based architectures, ARM-based CPUs \citep{McIntosh-Smith2019}, Field Programmable Gate Arrays (FPGAs), and many-core accelerator cards such as the Intel Xeon Phi have all been cited as possible architectures of the future \citep{Bergman}. As the recent discontinuation of the Intel Xeon Phi product line \citep{AnandTech} has highlighted however, it is not clear at this stage which, if any, of these architectures will ultimately prevail in the coming decades. As such, codes using cross-platform standardised libraries agnostic to the computational architecture, may be better placed to respond to future changes in the computational landscape. 

This section provides brief examples of OPS code that was automatically generated by OpenSBLI. Full descriptions of the OPS abstraction have been reported in \cite{OPSC,Reguly_2014_OPSC}. Unlike some black-box solvers, the OPS C code can be modified directly by the user. As a result, modifications can be made at either the Python, or C level, depending on the situation. New features will often be tested first in C, before being incorporated into the Python code-generation framework.

\begin{C}[caption={An example of calling a parallel loop region in OPS},captionpos=b]
// Example OPS parallel loop
int iteration_range_43_block0[] = {0, block0np0, 0, block0np1, 0, block0np2};
ops_par_loop(opensbliblock00Kernel043, "Convective CD p_B0 x1", opensbliblock00, 3, iteration_range_43_block0,
ops_arg_dat(p_B0, 1, stencil_0_02, "double", OPS_READ),
ops_arg_dat(wk20_B0, 1, stencil_0_00, "double", OPS_WRITE));
\end{C}

The core component of OPS is the \texttt{ops\_par\_loop} shown in the code example. This function provides a template to create a parallel region in OPS, based on the minimum amount of required information. Referring once more to figure \ref{fig:opensbli_flow_chart}, these templates are parsed by the OPS translator before compilation. The OPS translator is a Python script that reads and translates the base code to a range of parallel programming languages. For each individual \texttt{Kernel}, the computation is wrapped into parallel versions for each of the computational back-ends. The user gives OPS full control of the data, to decompose and execute the data most efficiently for a given platform. The example shown here is a kernel to perform a central derivative of $\partial p / \partial y$ at \nth{4} order.

In line 2 we specify the iteration range to be the entire $\left(N_x, N_y, N_z\right)$ grid. Line 3 begins the call to the parallel loop with an input function \texttt{Kernel043}, that contains the calculations to perform. The next arguments are a name for the computation, an OPS block that holds grid information, the dimensions of the problem, and the iteration range. Lines 4 and 5 specify the input/output arrays which are of type `\texttt{ops\_dat}'. We must provide the names, floating point precision, and access patterns for each of the arrays. Additionally, each array has a corresponding \texttt{stencil} argument, which is a list of integers for the relative stencil access. All of this information is generated automatically by the OpenSBLI \texttt{Kernel} class discussed in section \ref{sec:opensbli_kernels}. 


\begin{C}[caption={A user kernel to calculate a central derivative in OPS},captionpos=b]
// Example OPS kernel function
void opensbliblock00Kernel043(const double *p_B0, double *wk20_B0){
wk20_B0[OPS_ACC1(0,0,0)] = inv_1*(-rc7*p_B0[OPS_ACC0(0,2,0)] - rc8*p_B0[OPS_ACC0(0,-1,0)] + (rc8)*p_B0[OPS_ACC0(0,1,0)] + (rc7)*p_B0[OPS_ACC0(0,-2,0)]);}
\end{C}

The second example is the corresponding C function that was called within the \texttt{ops\_par\_loop}. The first thing to note here is the relative indexing for the computation. The code loops over the entire grid range in parallel, where the $\left(0,0,0\right)$ location refers to the current grid point. The pressure array `$p\_B0$' is being indexed in four points in the $y$ direction, to build the finite-difference approximation. The result is stored to the generic work array `$wk20\_B0$'. Work arrays are re-used where possible during the algorithm, to reduce the memory requirements of the code. The \texttt{OPS\_ACCX} macros correspond to the ordering of the input arguments to the function. These are another feature of OPS that would have to be maintained manually without code-generation. The code-generator also extracts all rational constant factors ($p/q$ for $p, q \in \mathbb{Z}$) found in the computational kernels, and defines them as global placeholder constants in the preamble of the simulation code. Having explained the design of OpenSBLI and the resulting OPS code, the next section shows a selection of test simulations.

\section{Validation and verification}\label{sec:results}
In this section, a selection of test cases are presented to demonstrate the capabilities of the OpenSBLI code for different flow configurations. The code repository contains other applications that are not discussed in detail here, as they have been covered in previous work. These applications include supersonic laminar SBLI \cite{LUSHER201817}, laminar duct SBLI with sidewalls \cite{Lusher2020_JFM}, transitional duct SBLI with sidewalls \cite{Lusher2020_FTAC}, and hypersonic roughness-induced transition at Mach 6 \cite{Lefieux2019}. The numerical schemes in OpenSBLI have also been assessed and validated for supersonic Taylor-Green vortex cases \cite{Lusher2019_AIAA}, and supersonic turbulent channel flows \cite{Arash_IFNMF}.

A set of verification and validation cases are documented in this section for problems that test specific parts of the code implementation. The Sod shock-tube problem in section \ref{sec:sod} shows the correctness of the shock-capturing and characteristic decomposition for a simple 1D case. The Shu-Osher shock-density wave interaction in section \ref{sec:shu_osher} highlights the benefits of high-order WENO shock-capturing schemes, in the context of resolving high-frequency waves. The order of convergence for the WENO-Z schemes is shown for a smooth propagating density wave in section \ref{sec:euler_wave}. This case is performed on a sinusoidal mesh to verify the curvilinear coordinate transformation against an exact solution. Section \ref{sec:laminar_channel} verifies the central and one-sided boundary schemes against an analytic solution for a 2D compressible laminar channel flow, for which a channel with one isothermal wall and one adiabatic wall are selected to concisely verify two no-slip wall conditions in the same problem. The 2D viscous shock-tube in section \ref{sec:viscous_shock_tube} highlights the ability of the code to capture a propagating normal shockwave, with subsequent vortex roll-ups and unsteady flow separation. Comparison is made to a reference solution at $Re=200$. The third order explicit Runge-Kutta time-stepping scheme is used for all test cases. The following section \ref{sec:transitional_SBLI} demonstrates a larger 3D SBLI DNS that contains acoustic-source body forcing, shock-reflection, flow-separation, and shock-induced transition to turbulence. 

\subsection{Sod shock-tube}\label{sec:sod}
\begin{figure}[!h]
\begin{center}
\includegraphics[width=1.0\columnwidth]{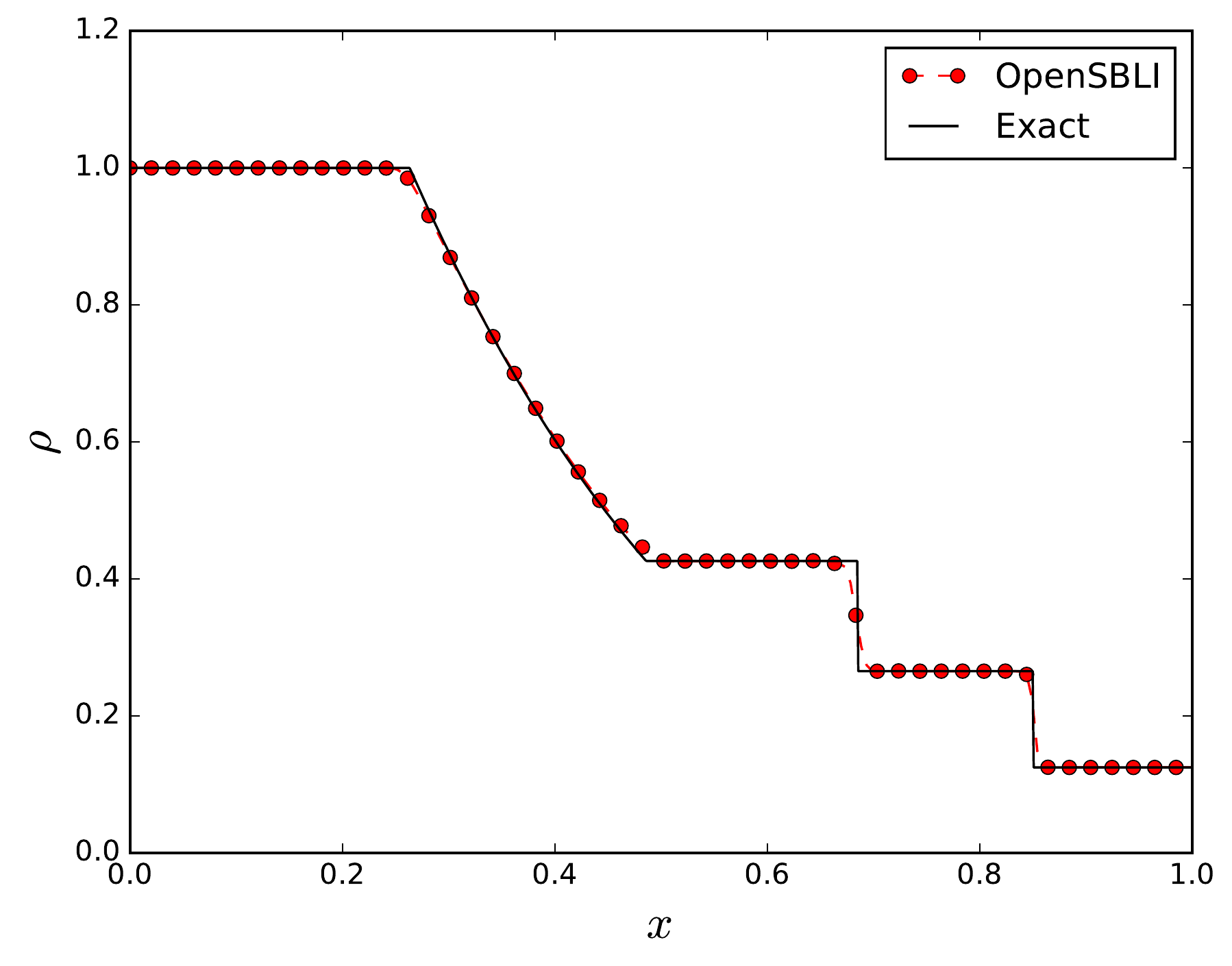}
\caption{Density profile for the Sod shock-tube case on $N=200$ grid points. Comparison is made to an exact solution, plotting every other point.}
\label{fig:sod_shocktube}
\end{center}
\end{figure}

The Sod shock-tube \cite{sod1978survey} is a classic test of shock-capturing ability for an ideal gas in one dimension. The test case consists of a Riemann problem specified by the initial left and right states separated at $x=0.5$ by
\[
\left(\rho, u, p \right) = 
\begin{cases}
    \left(1.0, 0.0, 1.0\right )  & \text{if } x < 0.5 \\
    \left( 0.125, 0.0, 0.1 \right) & \text{otherwise.}\\
\end{cases}
\]
The test acts as validation of the characteristic decomposition implementation in OpenSBLI. The initial state is advanced to a non-dimensional time of $t=0.2$, with constant time-step $\Delta t = 1\times 10^{-4}$. The TENO-5 scheme is selected with $N=200$ uniformly-spaced grid points. Figure \ref{fig:sod_shocktube} shows good agreement to the exact solution. There is an expected smearing of the steep discontinuities by the flux-splitting method at this resolution, similar to previous studies performed with these schemes \cite{HTR_code}. 

\subsection{Shu-Osher shock-density wave interaction}\label{sec:shu_osher}

\begin{figure}[!h]
  \center
  \includegraphics[width=1.0\columnwidth]{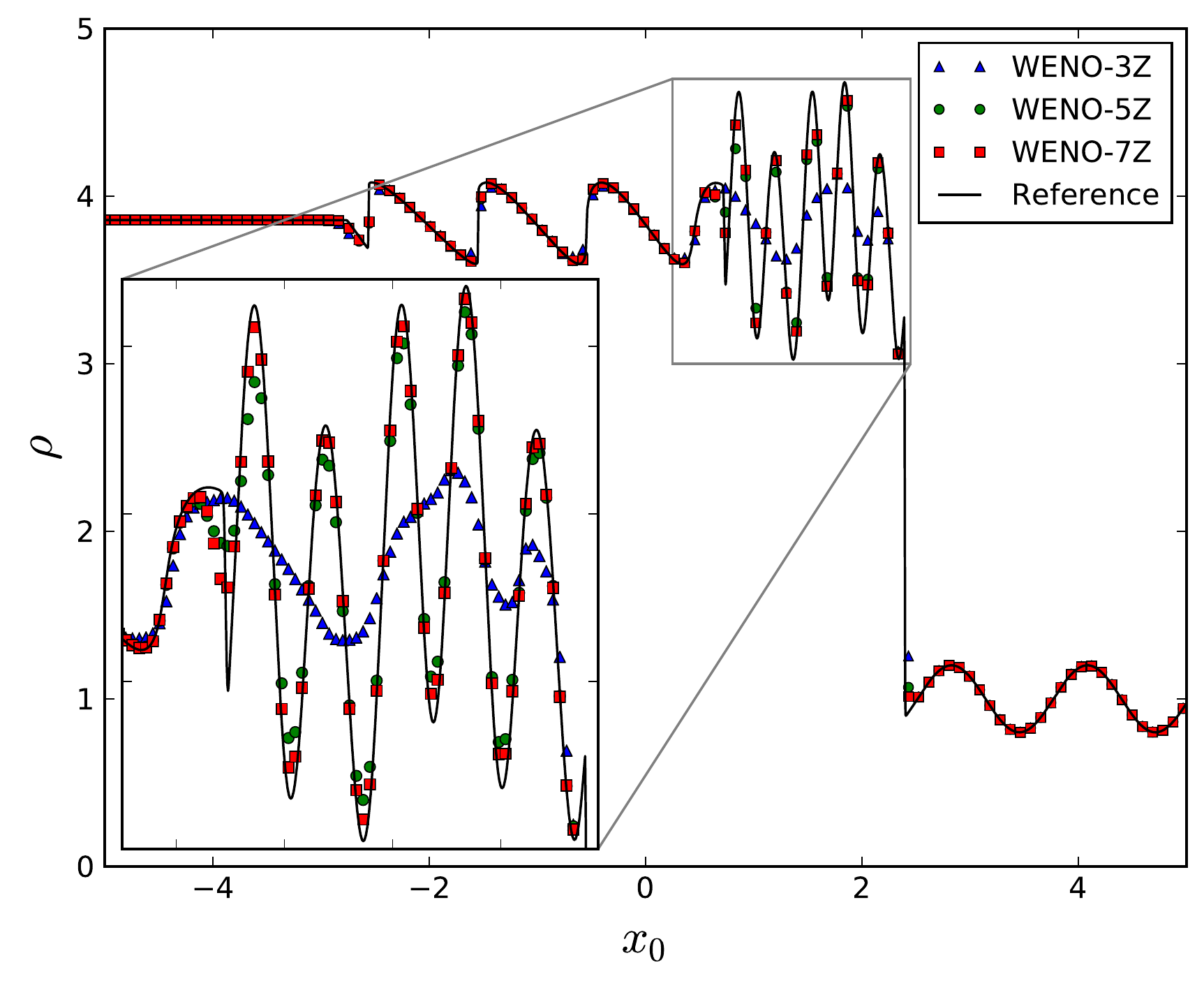}
  \caption{Shu-Osher density distribution at a non-dimensional simulation time of $t=1.8$, for WENO orders: 3Z (blue), 5Z (green) and 7Z (red). Reference fine mesh solution (continuous, black). Every third data point is displayed, with consecutive data shown in the 2x zoom inset. \label{fig:ShuOsher}}
\end{figure}

The Shu-Osher problem is a one-dimensional inviscid test case involving the interaction of a Mach 3 shock with a smooth density wave. Three different orders of the WENO-Z scheme are used in this section to highlight the benefits of higher-order shock-capturing schemes. The simulation is initialized with the discontinuous conditions given in \cite{Shu1988,Taylor2007}, such that 

\[
\left(\rho, u, p \right) = 
\begin{cases}
    \left( 3.857143, 2.629369, 10.33333 \right) & \text{if } x < -4 \\
    \left( 1+0.2\sin\left(5x\right), 1.0, 0.0 \right) & \text{if } x \geq -4\\
\end{cases}
\]
with Dirichlet conditions enforced at the domain boundaries $x = \left[-5, 5\right]$. The simulation is advanced to a non-dimensional time of $t=1.8$, with a time-step of $\Delta t = 2 \times 10^{-4}$. A reference solution was computed with a WENO-7Z scheme and a fine mesh of $N=3200$ grid points. Figure \ref{fig:ShuOsher} shows a comparison of results for the WENO-3Z, WENO-5Z, and WENO-7Z schemes on an $N=320$ grid. 

The normal-shock propagates in the positive $x$ direction from its start location at $x=-4$, interacting with the smooth imposed initial density perturbation. Shocklets are formed from the interaction in the region of $-2 < x < 1$, with a series of high-frequency waves present behind the normal-shock at $0 < x < 2$. The qualitative features of the solution are consistent with previous work such as figure 4 of \cite{Taylor2007}. While all three of the WENO-Z schemes match the reference solution for the shocklets and smooth regions of the flow, the lowest order WENO-3Z scheme struggles to resolve the high-frequency waves behind the main shock. The zoomed-inset highlights the benefits of using higher-order schemes.

\subsection{Curvilinear Euler 2D-wave propagation}\label{sec:euler_wave}
\begin{figure}[!ht]
\begin{center}
\includegraphics[width=1.0\columnwidth]{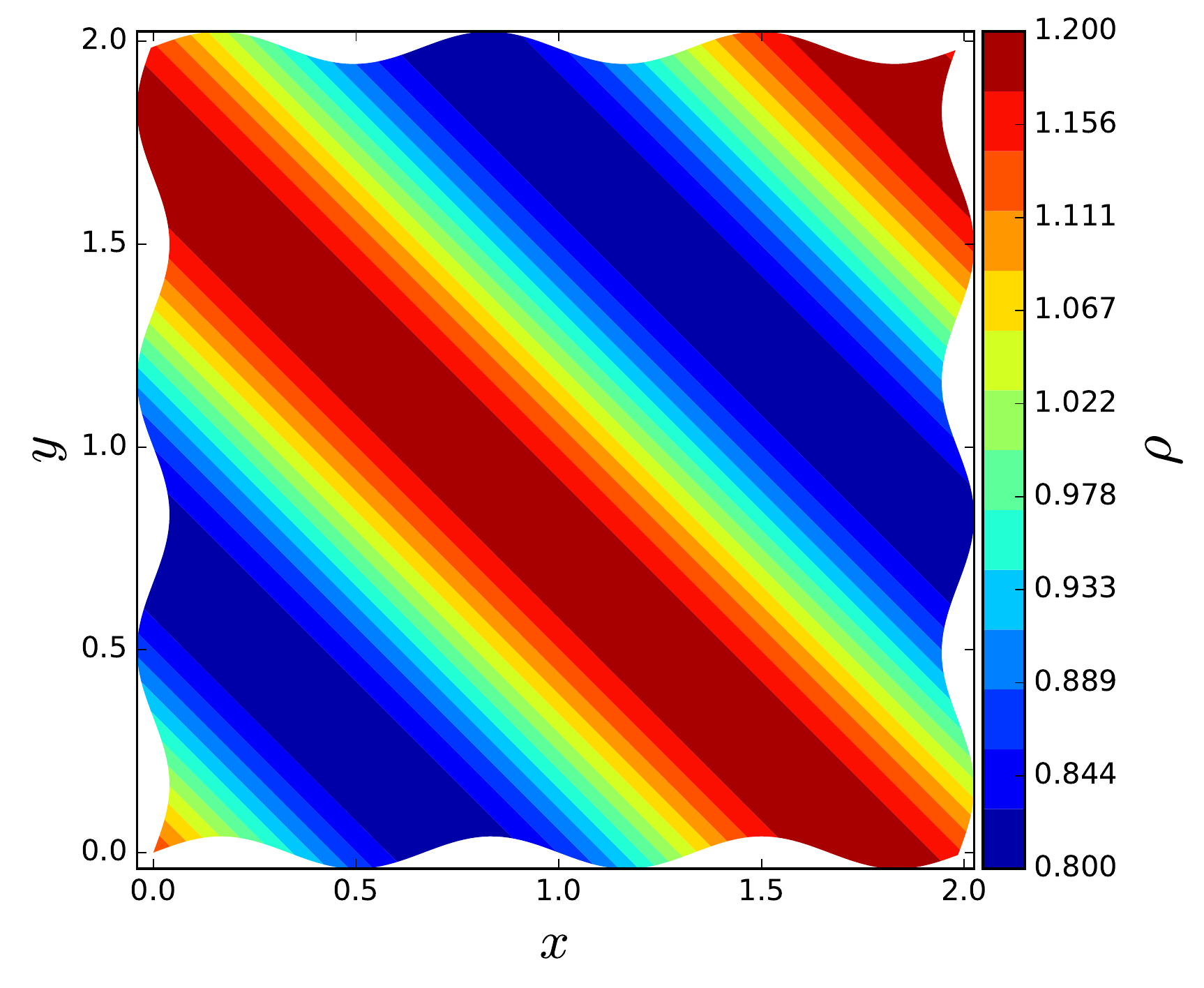}
\caption{Travelling 2D density wave, for the Euler equations simulated on a sinusoidal curvilinear mesh \eqref{curved_grid}.}
\label{fig:Euler_contours}
\end{center}
\end{figure}

WENO and TENO schemes are adaptive in the sense that the number of candidate stencils (figure \ref{fig:weno_stencil}) used for the flux reconstruction will vary depending on the local flow conditions. The schemes will reduce to lower order approximations around shocks, and maintain the full numerical stencil in smooth regions of the flow. To be able to test the formal order of convergence of the schemes, it is therefore necessary to apply them to a smooth flow. The selected case is a smooth travelling 2D density perturbation for the Euler equations. The initial perturbation takes the form
\begin{equation}\label{euler_wave}
\rho(x,y,t) = 1 + 0.2\sin\left(\pi\left(x+y - t\left(u+v \right)\right)\right),
\end{equation}
with constant velocity components $u = 1.0$, $v = -0.5$, and a pressure of $p=1.0$. Periodic boundaries are applied on all sides of the domain, with a time-step of $\Delta t=5\times 10^{-4}$. To verify the implementation of the Euler equations and characteristic decomposition on curvilinear meshes, a comparison is made to the analytical solution after a non-dimensional time of $t=2.5$. A sinusoidal grid is generated such that
\begin{align}\label{curved_grid}
x_{i,j} &= i\Delta x + A\sin\left(\frac{6\pi j \Delta y}{L_y}\right),\\
y_{i,j} &= j\Delta y + A\sin\left(\frac{6\pi i \Delta x}{L_x}\right),
\end{align}
for $A=0.04$, and $L_x = 2/L_x$, $L_y = 2/N_y$. The solution is shown on the curvilinear mesh in figure \ref{fig:Euler_contours}.

At each grid resolution, the $L^{1}$ and $L^{\infty}$ error norms are computed as
\begin{align}
L^{1} &= \frac{\sum_{i,j} \left|\rho_{\textrm{exact}} - \rho_{i,j}\right|}{N_{x}N_{y}},\\
L^{\infty} &= \left|\rho_{\textrm{exact}} - \rho_{i,j}\right|_{\rm{max}}.
\end{align}
\begin{figure}[!ht]
\begin{center}
\includegraphics[width=1.0\columnwidth]{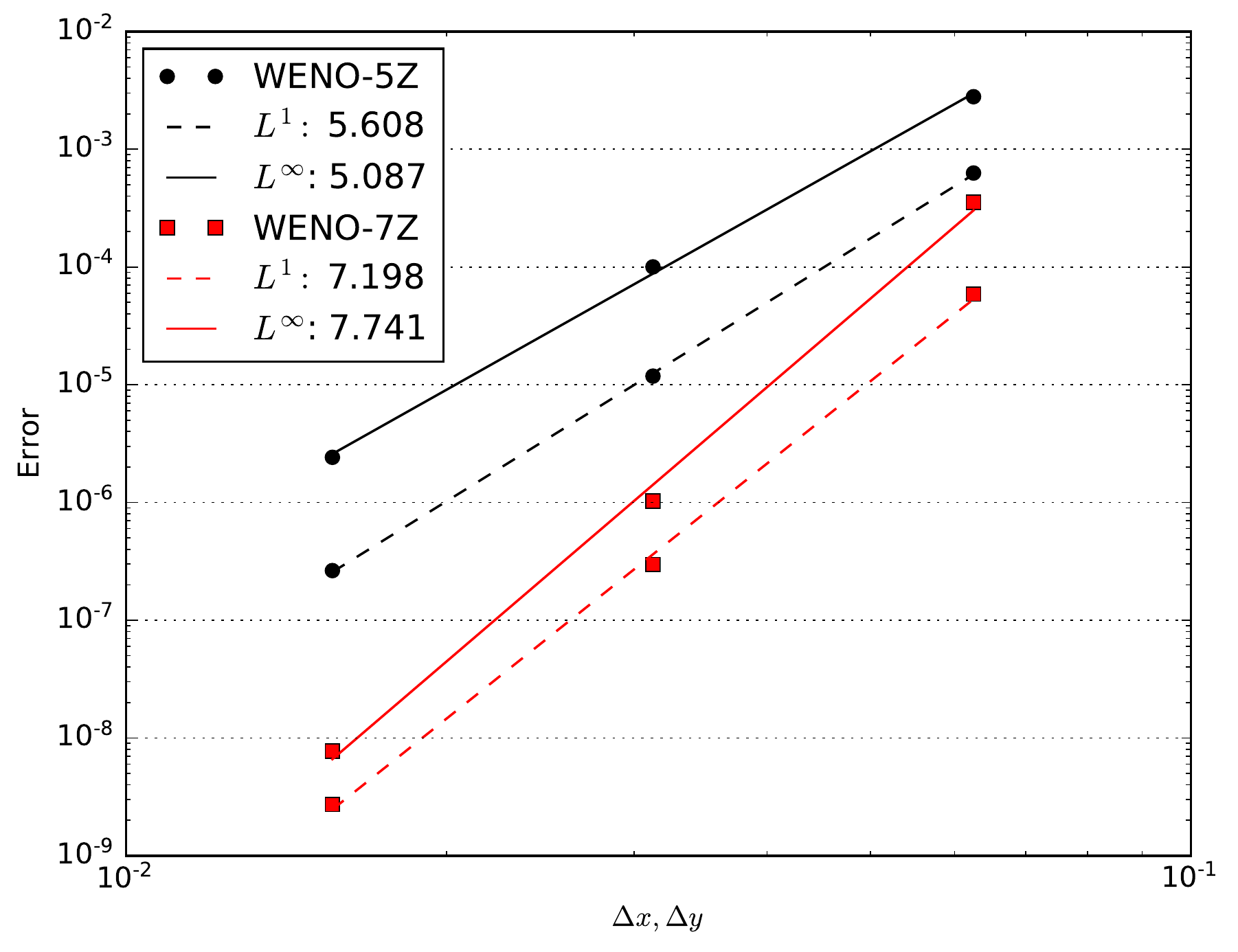}
\caption{Convergence of the $L^{1}$ and $L^{\infty}$ error norms for WENO-Z schemes on a curvilinear mesh. The grid sizes are $N=\left[32^2, 64^2, 128^2\right]$.}
\label{fig:euler_wave_error}
\end{center}
\end{figure}
Figure \ref{fig:euler_wave_error} shows the rate of convergence for WENO-Z schemes of \nth{5} and \nth{7} order on $N=\left[32^2, 64^2, 128^2\right]$ grids. In both cases the schemes converge towards the exact solution with grid refinement, verifying the implementation of the curvilinear coordinate transformation. Both the $L^{1}$ and $L^{\infty}$ convergence rates are within 10\% of the formal order of accuracy for these shock-capturing schemes.

\subsection{Laminar compressible channel with mixed adiabatic/isothermal wall boundaries}\label{sec:laminar_channel}

To verify the implementation of the isothermal and adiabatic no-slip wall boundary conditions, a compressible laminar channel flow is compared to an analytical solution. The channel is comprised of a constant temperature wall at $y=-1$, and a zero heat-flux wall at $y=1$. For constant viscosity, the streamwise velocity and temperature profiles for the analytic solution are given by

\begin{align}\label{laminar_channel_analytic}
        u &= \frac{Re(1-y^2)}{2}, \\
        T &= 1 - \frac{{Re}^2 M_{\rm{ref}}^{2} Pr(\gamma-1) (y^4-4y-5)}{12}.
\end{align}

Simulation parameters are taken to be $Re=90$, $Pr=0.72$, and $M_{\rm{ref}} = 0.01$. The isothermal wall temperature is set to 1. At these physical values, the analytical solution predicts an adiabatic wall temperature at $y=1$ of $T_{aw} = 1.155520$ (7 s.f.). The simulation is advanced until a non-dimensional time of $t=1000$, with a time-step of $\Delta t = 1\times 10 ^{-4}$. This test case was run as a 2D problem in a domain with size $L_x=2\pi$, $L_y=2$, with a uniform grid distribution of $\left(N_x, N_y\right) = \left(32, 64\right)$.

\begin{figure}[!ht]
\begin{center}
\includegraphics[width=1.0\columnwidth]{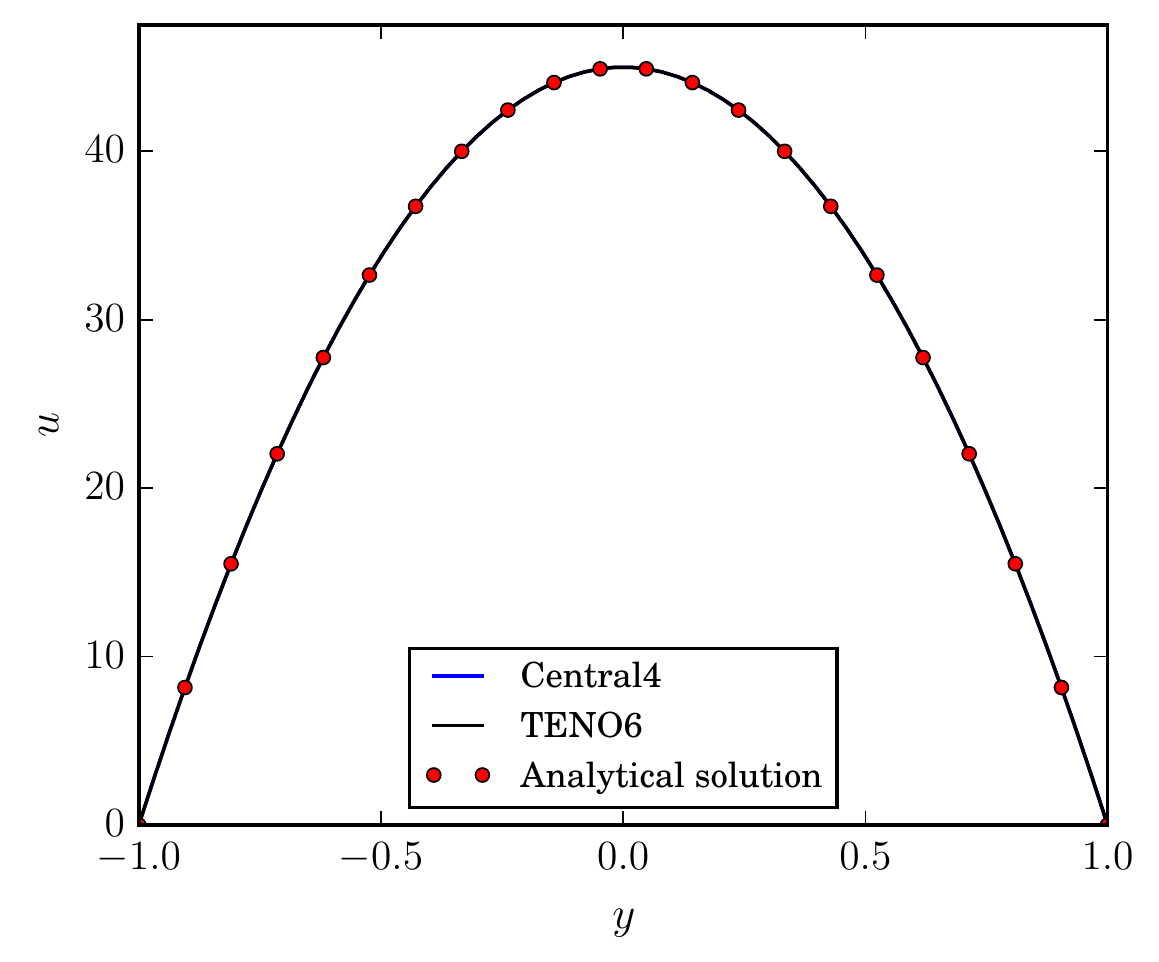}
\includegraphics[width=1.0\columnwidth]{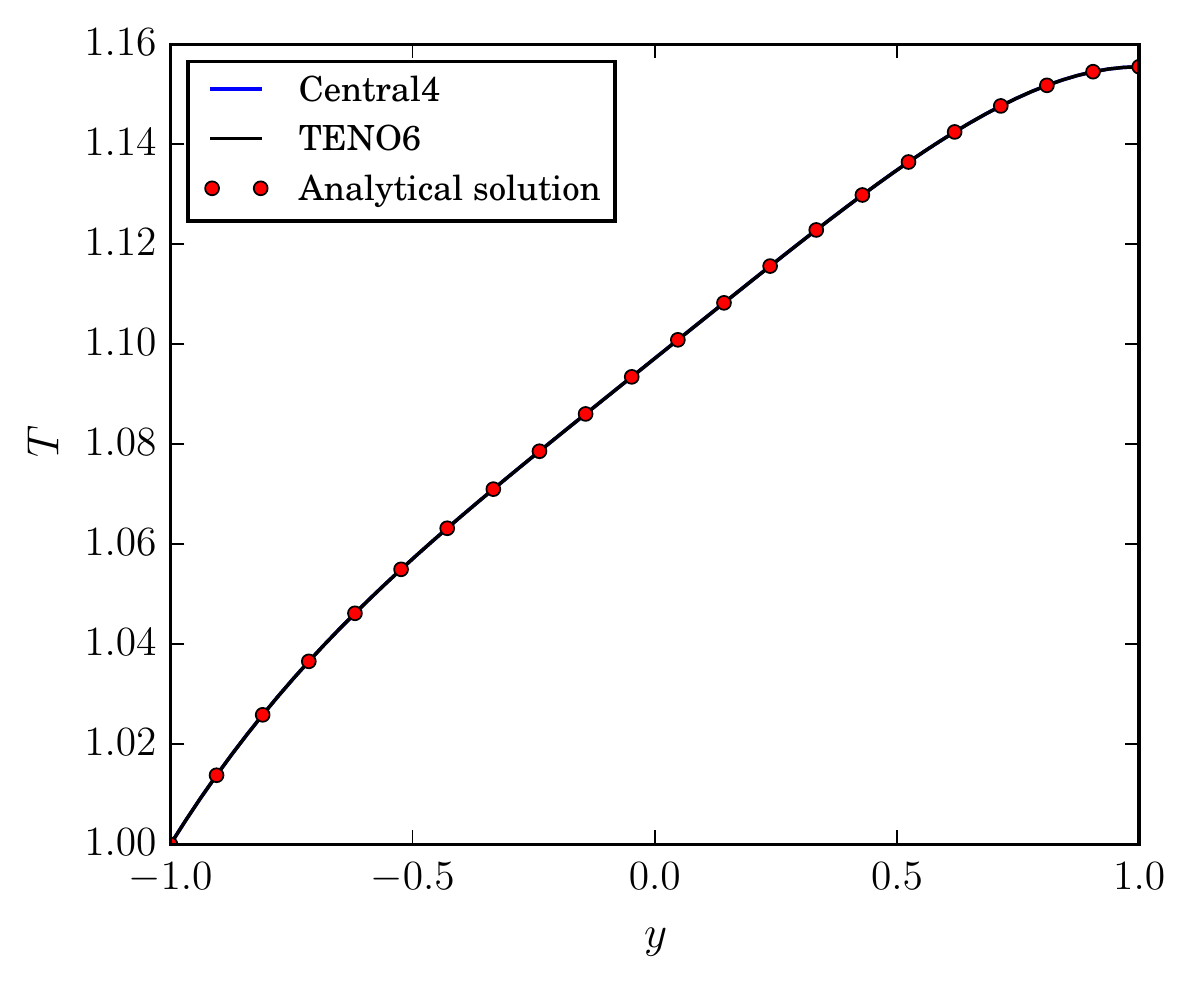}
\caption{Comparison of the 2D mixed-wall condition laminar channel flow to the analytical solution in equation \eqref{laminar_channel_analytic}. Showing profiles for the (top) streamwise velocity and (bottom) temperature.}
\label{fig:laminar_channel_results}
\end{center}
\end{figure}

Figure \ref{fig:laminar_channel_results} shows the results for both a \nth{4} order central scheme and a \nth{6} order TENO scheme. Excellent agreement is observed to the analytical result for both the streamwise velocity and temperature profiles. The correct near-wall behaviour is produced for both of the schemes. At this grid resolution, the central and TENO adiabatic wall temperatures are 1.155510 and 1.155508 respectively. This corresponds to relative percentage errors for $T_{aw}$ of $0.8\times 10^{-3}$ \% and $1.0\times 10^{-3}$ \% for the central and TENO schemes respectively. 

\subsection{2D Viscous shock-tube}\label{sec:viscous_shock_tube}
\begin{figure}[!ht]
\begin{center}
\includegraphics[width=1.0\columnwidth]{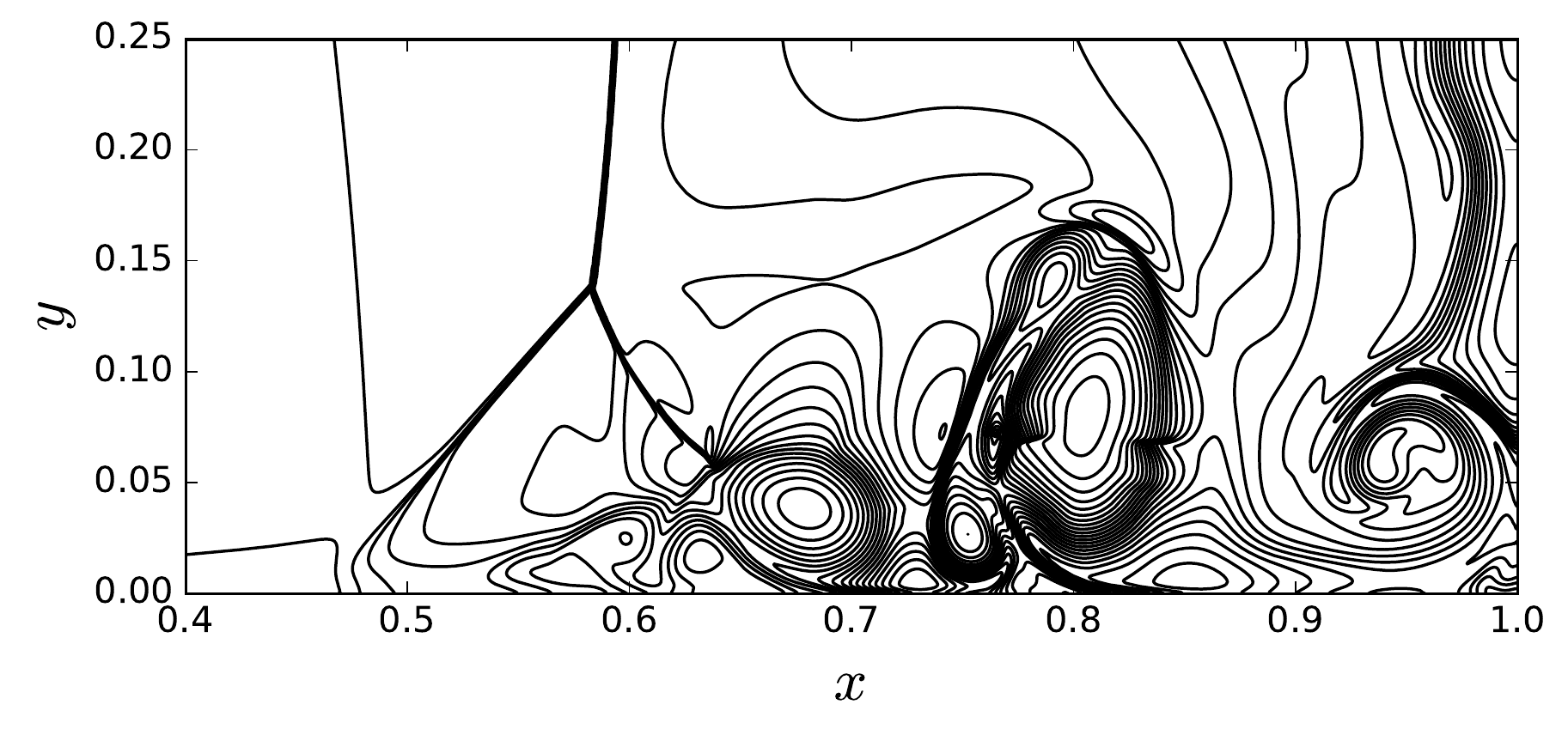}
\caption{Instantaneous density contours for the viscous shock-tube problem at $Re=200$, on a $\left(N_x, N_y\right) = \left(1500, 750\right)$ mesh at a time of $t=1$.}
\label{fig:viscous_shocktube_contours}
\end{center}
\end{figure}

\begin{figure}[!ht]
\begin{center}
\includegraphics[width=1.0\columnwidth]{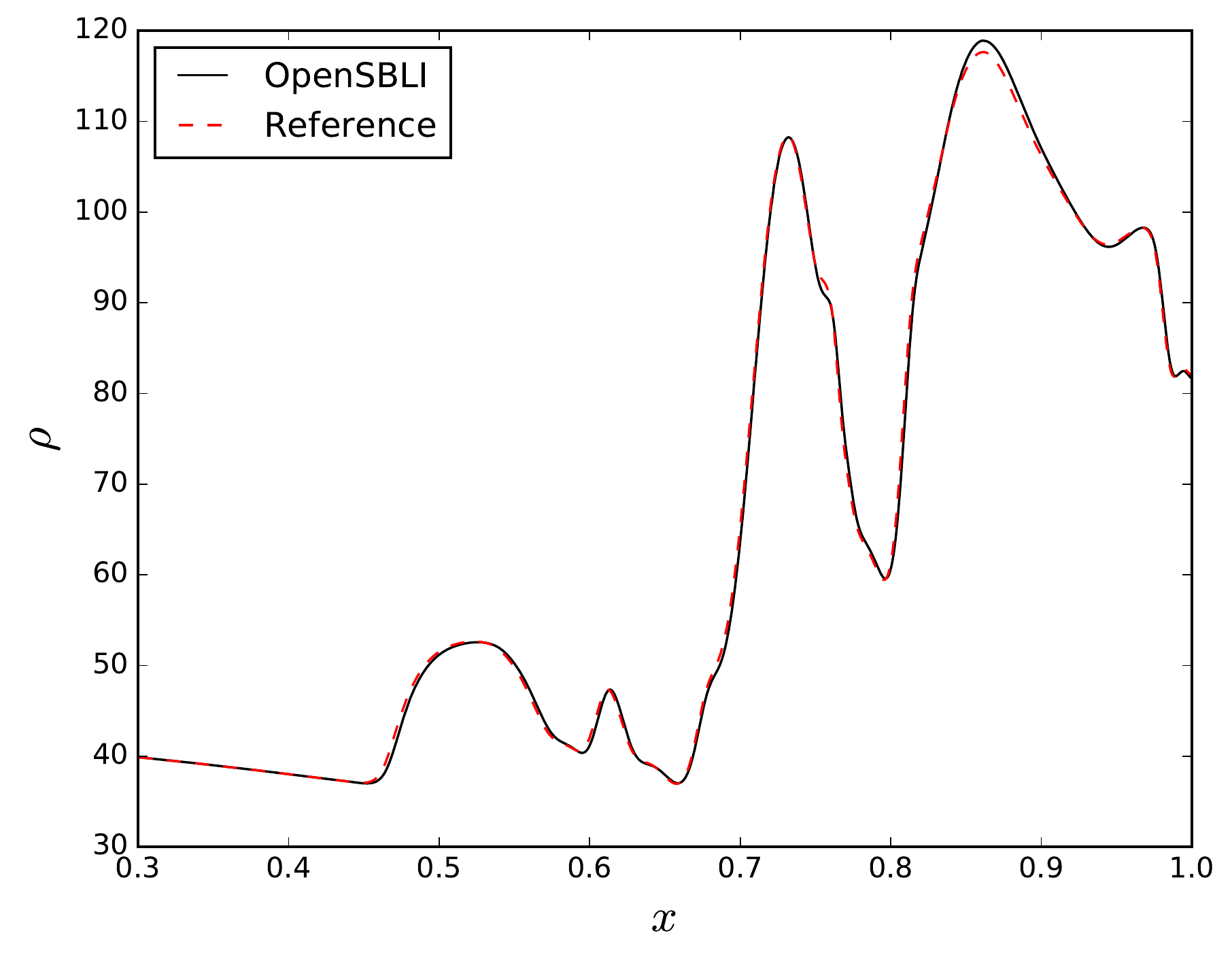}
\caption{Wall density for the viscous shock-tube problem at $Re=200$, on a $\left(N_x, N_y\right) = \left(1500, 750\right)$ mesh. Compared to a reference solution \cite{Zhou2018} at a time of $t=1$.}
\label{fig:viscous_shocktube_line}
\end{center}
\end{figure}

The viscous shock-tube is a demanding case for shock-capturing schemes and tests the ability of the code to simulate a propagating shockwave interacting with a no-slip wall. The problem combines complex shock structures, shock reflection from a solid wall, and regions of unsteady boundary-layer separation. An in-depth study of the inviscid and viscous 2D shock-tubes was given by \cite{Daru2009}, from which the flow conditions in this section are taken. The computational domain $(x, y) = ([0,1], [0,0.5])$ is partitioned by an initial diaphragm located at $x=0.5$. A discontinuous initial profile is imposed which generates a normal shock that propagates in the positive $x$ direction. The initial states to the left and right of the diaphragm are given by
\[
\left(\rho, u, v, p \right) = 
\begin{cases}
    \left( 120, 0, 0, 120/\gamma\right) & \text{if } x < 0.5 \\
    \left( 1.2, 0, 0, 1.2/\gamma \right) & \text{if } x \geq 0.5.\\
\end{cases}
\]
The reference Mach number is set to $M_{\rm{ref}}=1$, with Reynolds number $Re= 200$, and a Prandtl number of $Pr = 0.73$. The viscosity is assumed to be constant. A symmetry condition is enforced on the upper boundary of the domain, with adiabatic no-slip viscous wall conditions set on all other boundaries. The simulation is advanced with time-step of $\Delta t = 1\times10^{-5}$ until a non-dimensional time of $t=1$. The grid consists of a uniform distribution of $\left[x, y\right] = \left[1500, 750\right]$ points, following the resolution of \cite{Zhou2018}.

Figure \ref{fig:viscous_shocktube_contours} shows the instantaneous density contours at $t=1$. We observe that the shock has propagated to the $x=1$ adiabatic wall, and reflected back to a position of $x=0.58$. The relative motion of the shock generates a thin boundary-layer which separates after the shock reflects from the side wall. Above the separation bubble a lambda shock structure can be seen to form. Good agreement is found to the results of \cite{Zhou2018} (figure 3), with the lambda-shock triple point located at $\left(x, y\right) \left(0.58, 0.13\right)$. Figure \ref{fig:viscous_shocktube_line} shows a comparison of the bottom wall density profile compared to the reference data of \cite{Zhou2018}. Good agreement is observed relative to the reference solution, aside from a small overshoot in the density peak located at $x=0.85$. 

\section{3D transitional shockwave/boundary-layer interaction}\label{sec:transitional_SBLI}
\begin{figure}
  \center
  \includegraphics[width=1.0\columnwidth]{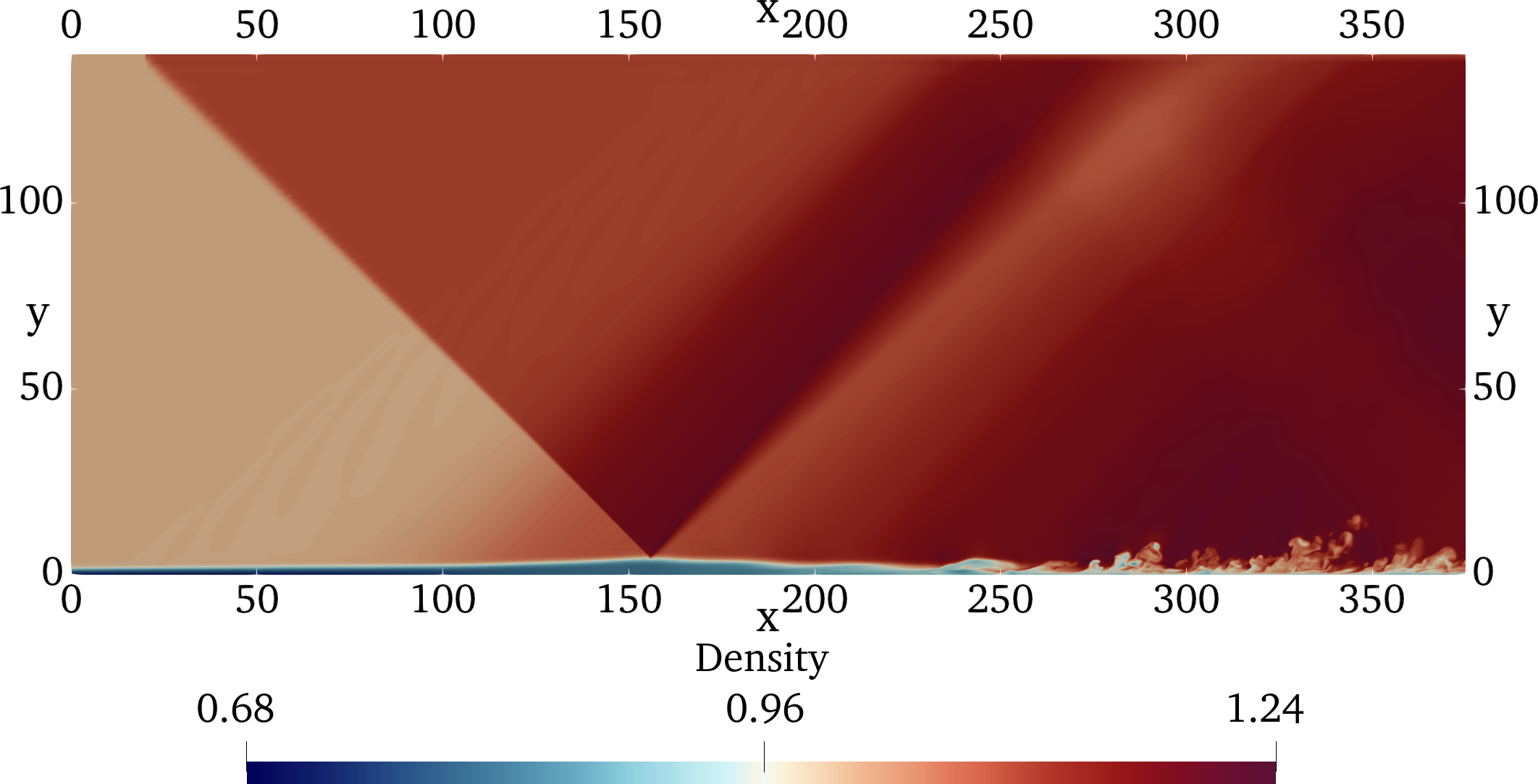}
  \caption{An $x$-$y$ slice of instantaneous density for the transitional shockwave/boundary-layer interaction. \label{fig:SBLI_density}}
\end{figure}

As a demonstration of a complex large-scale computation, a DNS of a Mach 1.5 transitional SBLI is performed. The case consists of a modified version of the simulations presented in \cite{sansica_thesis}. Compared to the original case, the domain is elongated in the streamwise direction to observe more of the breakdown process, and a different off-wall forcing method is used to seed the instability. 

A Mach 1.5 freestream is initialized throughout the domain, with the similarity solution for a laminar boundary-layer \cite{white2006viscous} in the near-wall region. A Reynolds number based on an inlet boundary-layer displacement thickness of $Re_{\delta^{*}} = 750$ is used, with a reference temperature of $T_{\infty} = 202.17K$. Details of the laminar boundary-layer similarity solution are given in \cite{LUSHER201817,Lusher2020_JFM}. The similarity solution generates profiles for temperature, streamwise velocity, and wall-normal velocity, which are then used to set the conservative variables throughout. The domain has dimensions of $L_x=375$, $L_y=140$, and $L_z=27.32$. Rankine–Hugoniot shock-jump conditions are enforced on the upper Dirichlet boundary at $x = 20$, corresponding to a flow deflection of $2.5^{\circ}$ and pressure ratio $p_2 / p_1 = 1.132$. Extrapolation methods are used at the inlet (for pressure) and outlet, while the span is periodic. Isothermal no-slip conditions are set on the bottom wall using the non-dimensional wall temperature of $T_{w} = 1.381$ (4 s.f.) from the similarity solution. 

\begin{figure}
  \center
  \includegraphics[width=1.0\columnwidth]{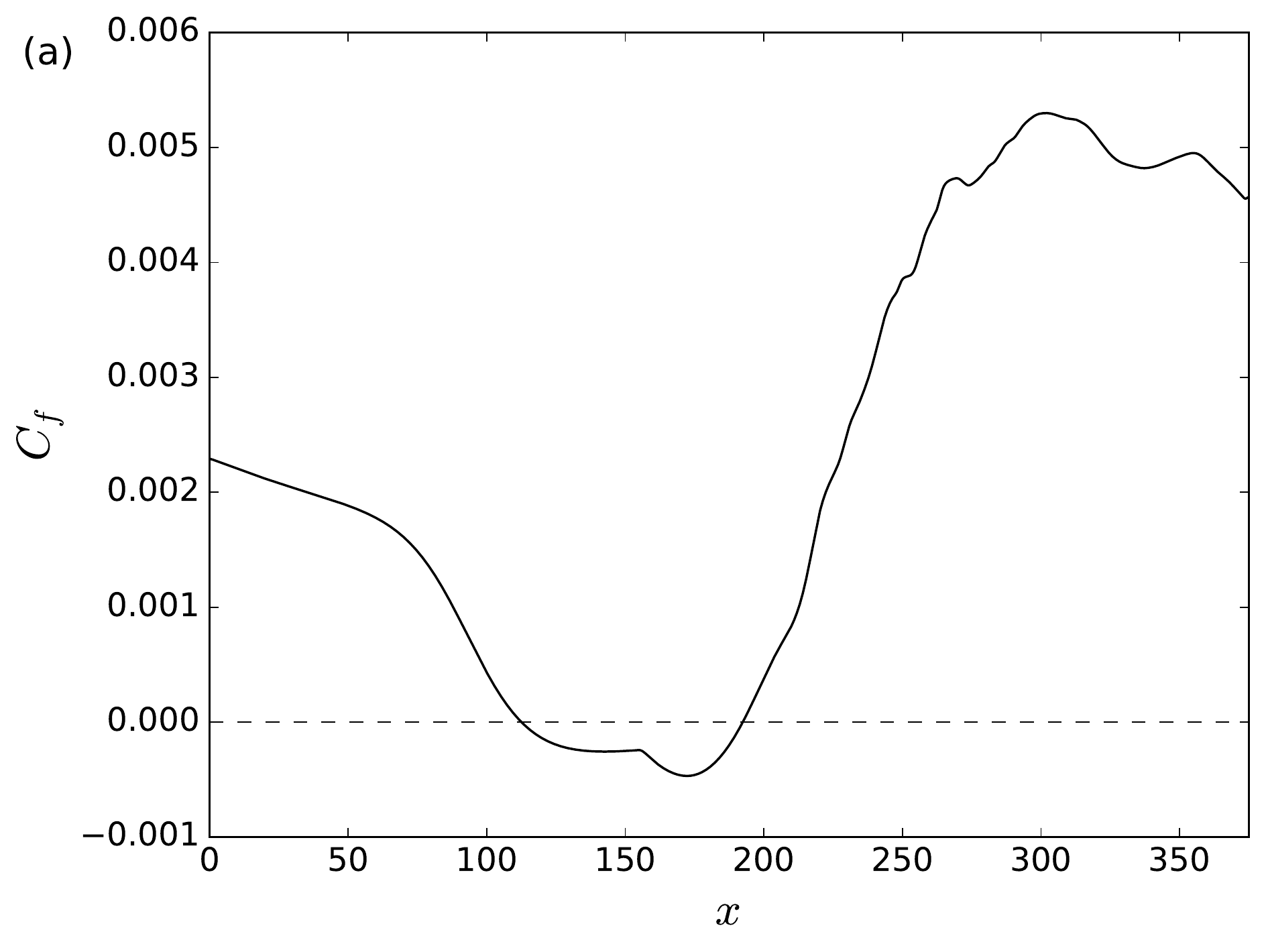}
  \includegraphics[width=1.0\columnwidth]{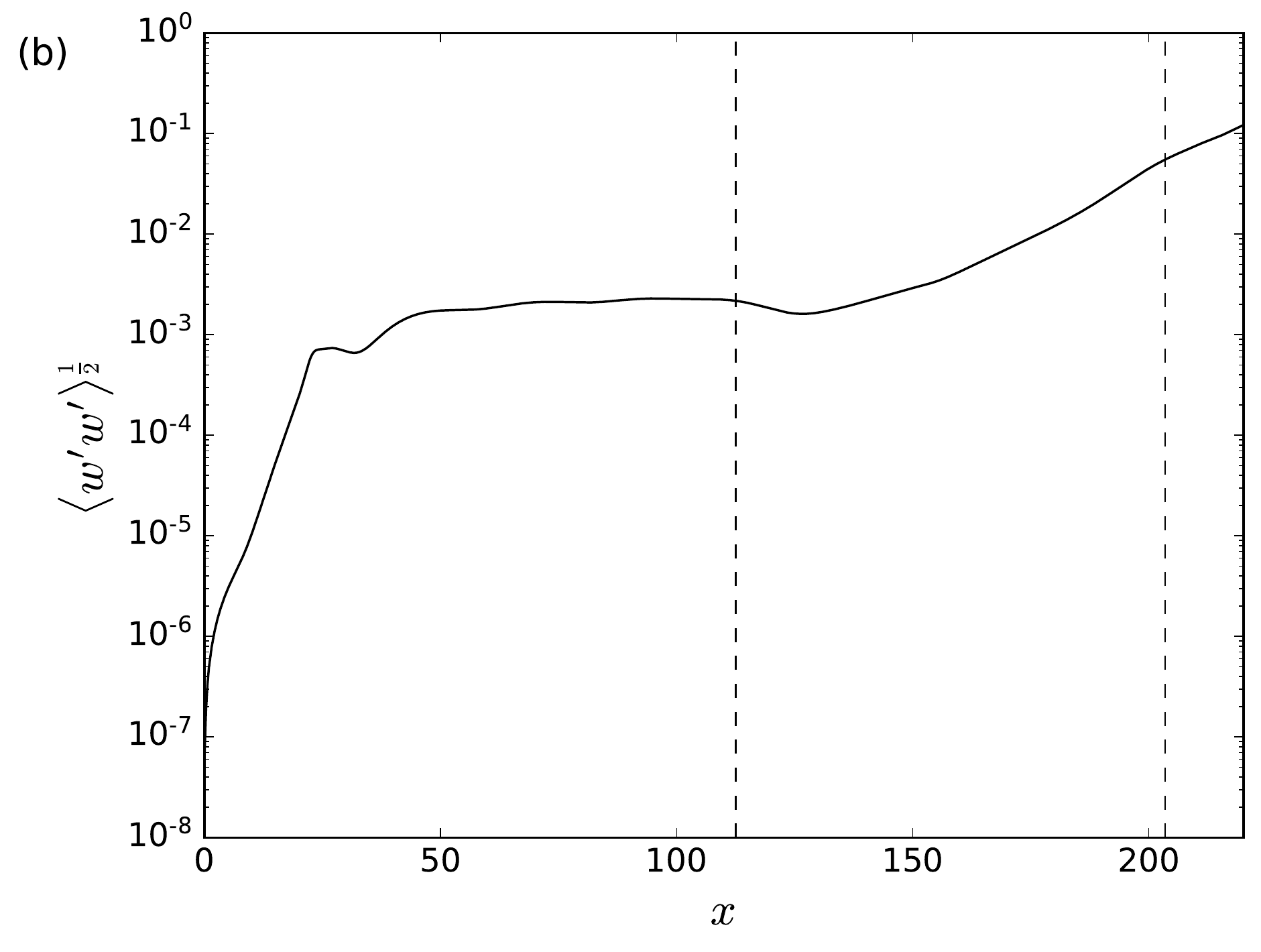}
  \caption{(a) Streamwise distribution of the time-averaged skin-friction coefficient for the 3D SBLI case. (b) Amplification of the $w$ velocity RMS of the disturbance along the line $y=1$, $z=L_z/4$. The dashed lines indicate the start and end of the separation bubble. \label{fig:SBLI_line}}
\end{figure}

For low-supersonic boundary-layers, it has been shown that the dominant transition mechanism is the breakdown of oblique first-mode waves \cite{FaselOblique, chang_malik_1994}. To introduce upstream disturbances to the otherwise laminar SBLI, modal time-dependent forcing is applied as an acoustic body-forcing term in the continuity equation \eqref{ns_eqn}. The forcing takes the form
\begin{equation}\label{SBLI_forcing}
\rho' = A \exp\left(-\left((x-x_F)^2+(y-y_F)^2\right)\right) \cos\left(\beta z\right) \sin\left(\omega t\right),
\end{equation}
for an amplitude $A=2.5\times 10^{-3}$, frequency $\omega = 0.1011$, and wavenumber $\beta = 0.23$. This single mode corresponds to the most unstable mode obtained from linear stability analysis of a laminar separation bubble \cite{sansica_thesis}. The wavenumber $\beta = 2\pi / L_z$, corresponds to one wavelength across the width of the span. The acoustic source is located in the freestream above the boundary-layer at $x_F=20$, $y_F=4$. The simulation was performed with the \nth{6} order TENO scheme, \nth{4} order central-differencing for diffusive terms, and \nth{3} order Runge-Kutta time-stepping. The number of grid points is taken to be $\left(N_x=2050, N_y=325, N_z=200\right)$. The grid is stretched in the streamwise and wall-normal directions such that 
\begin{align}
x_i &= L_x \left(1 - \frac{\sinh\left(s_x \Delta_x \left(N_x-1-i\right)\right) / L_x}{\sinh\left(s_x\right)}\right), \\
y_j &= L_y \frac{\sinh\left(s_y \Delta_y j / L_y\right)}{\sinh\left(s_y\right)},
\end{align}
for grid indices $i$, $j$, and stretch factors of $s_x=1.5$, and $s_y=5$. In wall units this corresponds to $\Delta x^{+} = 4.2$, $\Delta y^{+} = 0.95$, and $\Delta z^{+} = 4.5$, based on time-averaged skin-friction in the early turbulent region. A non-dimensional time-step of $\Delta t = 1\times 10^{-2}$ was used. The simulation was initially advanced in time for 7 flow-through times of the domain to allow the SBLI to develop. Statistics were gathered every iteration for a further 60 periods of the forcing.

Figure \ref{fig:SBLI_density} shows an ($x$-$y$) slice of instantaneous density contours, to highlight the main features of the shockwave/boundary-layer interaction. The initial oblique shockwave impinges on the flow developing over the wall, causing a thickening of the target boundary-layer. A series of compression waves are observed at $x=100$, as the supersonic freestream adjusts to the curvature of the boundary-layer. A separation bubble is present beneath the foot of the shock reflection. This feature is clearer to see in the time-averaged skin-friction in figure \ref{fig:SBLI_line} (a). Due to the adverse pressure gradient applied by the shockwave, the flow detaches from the wall at $x=115.5$, and reattaches at $x=192.4$. A transition to turbulence is observed in figure \ref{fig:SBLI_density}, downstream of the separation bubble. At $x=200$ in figure \ref{fig:SBLI_line} (a) there is a sharp increase in skin-friction, as the flow undergoes the early stages of transition. The skin-friction decreases close to the outlet, in the later stages of the turbulent breakdown. The exit skin-friction is approximately an order of magnitude higher than for a laminar version of the SBLI (e.g. figure 4.8 of \cite{sansica_thesis}).

\begin{figure}
  \center
  \includegraphics[width=1.0\columnwidth]{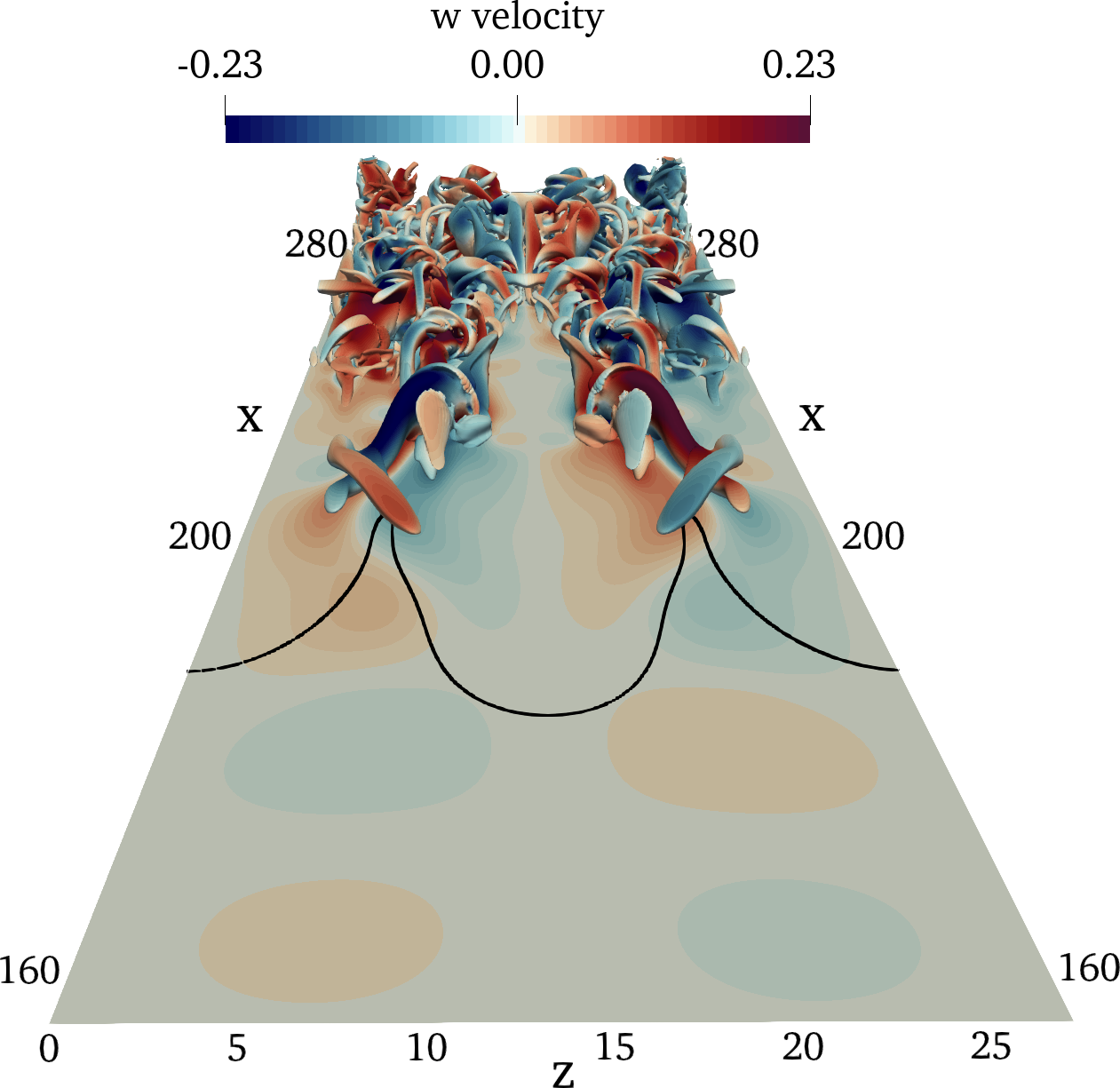}
  \caption{$Q$-criterion of $Q=5\times 10^{-3}$ coloured by $w$ velocity, showing the initial oblique mode breakdown of the transitional SBLI. A slice of $w$ velocity is shown at $y=0.5$, with the separation bubble ($u < 0$) in this plane outlined in black. \label{fig:SBLI_breakdown}}
\end{figure}

Figure \ref{fig:SBLI_line} (b) shows the RMS value of the $w$ velocity component induced by the freestream acoustic forcing. The disturbance is evaluated within the boundary-layer, along the line $y=1$, $z=L_z/4$. The two vertical dashed lines denote the start and end of the separation bubble on the line $z=L_z/4$. The SBLI amplifies the disturbance waves exponentially by two orders of magnitude, triggering a non-linear breakdown of the flow for $x > 210$. This feature is better visualized in the 3D-view of figure \ref{fig:SBLI_breakdown}, where vortical structures are shown for a Q-criterion of $Q=5\times 10^{-3}$, coloured by the instantaneous spanwise $w$ velocity. Beneath the $Q$ surfaces is a slice of $w$ velocity at $y=0.5$, with the black line representing the $u=0$ boundary of the separation bubble in this plane. A pair of symmetric streamwise vortices are present behind the reattachment line, which generate smaller vortex structures that remain symmetric for the range shown. Consistent with previous studies (e.g. figure 4.6 \cite{sansica_thesis}, figure 17 \cite{chang_malik_1994}), the dominant transition mechanism is observed to be an oblique-mode breakdown. This section has demonstrated that the code is capable of performing high-fidelity DNS of transitional SBLI problems, involving shock reflections, flow separation, and breakdown to turbulence.

\section{Performance and scaling}\label{sec:performance}
In this section, the performance and scaling of the OpenSBLI/OPS code is demonstrated on multi-GPU clusters. In particular, we highlight scaling differences between two of the main schemes in the current version of the code. The performance of the OPS library has been documented extensively in previous studies, over a wide range of computational platforms. These include comparison of each of the computational back-ends to hand-written code \cite{OPS_perf}, performance on ARM-based systems \cite{McIntosh-Smith2019}, and an in-depth comparison against a comparable structured-mesh DNS code \cite{Mudalige2019}. The results in this section use a standalone OpenSBLI performance benchmark configured for scaling tests. For consistency with previous studies, the benchmark is configured to perform DNS of the 3D Taylor-Green Vortex (TGV) problem. A full description of the TGV case has been given in the context of OpenSBLI in \cite{JACOBS201712, Lusher2019_AIAA, Mudalige2019}. The problem contains the transition and turbulent breakdown of an initial vortex condition on a triply periodic domain. 

\subsection{Low memory intensity algorithms}
\begin{figure}
  \center
  \includegraphics[width=1.0\columnwidth]{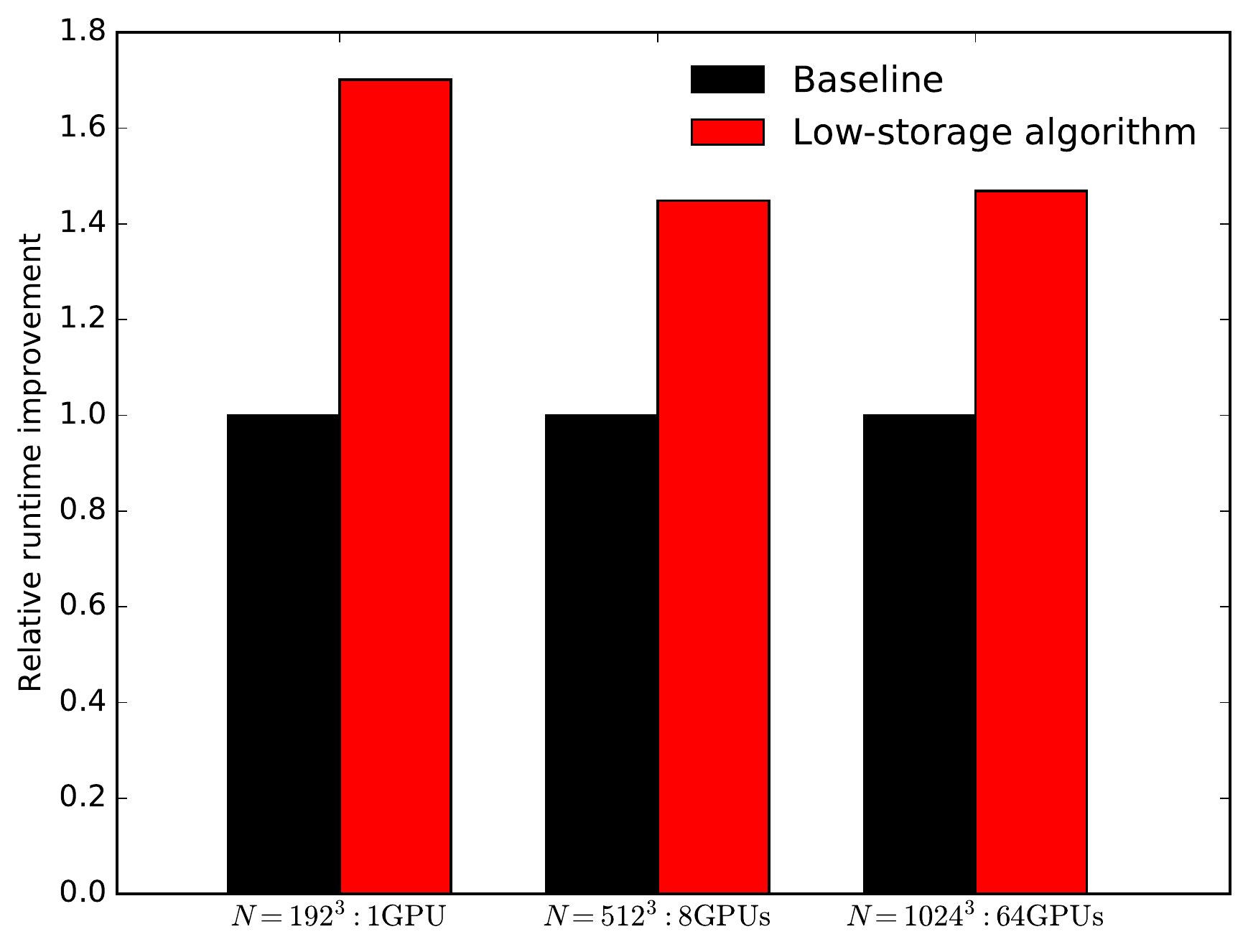}
  \caption{Relative runtime speed-up factor of the low-memory intensity central-differencing algorithm. Simulations are performed on $N=\left[192^3, 512^3, 1024^3\right]$ grids on 1, 8, and 64 GPUs respectively.}\label{fig:SS_performance}
\end{figure}

In recent decades, increases in available compute capability (FLOPS) have greatly outpaced increases in memory bandwidth \cite{Witherden2014}. This has led to many CFD codes being limited on performance by memory access. Minimising access to global memory by re-computing quantities locally on the fly is one way to alleviate this issue. Furthermore, high-order algorithms performing a large number of operations per-byte are well suited to modern hardware. For all of the simulations shown in this section, MPI+CUDA executables were generated using the OPS library discussed in section \ref{sec:OPS_section}. Run-times were compared for 100 iterations of the main time loop in each case, omitting the input/output time of the simulation. Scaling results were obtained for Nvidia P100 GPU partitions on the CSD3 high-performance-computing cluster at the University of Cambridge. CUDA version 10.1 was used with -O3 optimization, and the Intel compiler (2017) and OpenMPI for inter-GPU and inter-node MPI communication.

Code-generation enables manipulation of the equations and solution algorithm that would be difficult to achieve in a hand-written code. To demonstrate the benefit of reduced memory access, we show the \texttt{StoreSome} optimisation \cite{Jammy_Algo} that is available in OpenSBLI for the central schemes. The standard practice in CFD is to store derivatives globally on the entire grid, to be re-accessed at later stages in the algorithm. The \texttt{StoreSome} algorithm in contrast, computes most of the derivatives locally in the kernel (section \ref{sec:opensbli_kernels}). Based on previous work \cite{Jammy_Algo}, only the first derivatives of the velocity components ($u,v,w$) are stored in global arrays. Figure \ref{fig:SS_performance} shows the runtime improvement of the \texttt{StoreSome} algorithm. Compared to the baseline version, the low-memory intensity algorithm is 1.7x faster on a single GPU. The improvement drops to 1.45x on multi-GPU configurations, as part of the total runtime is now taken up by inter-node MPI exchanges over the CPU hosts. 

A secondary benefit of the algorithm that is especially pertinent on GPUs, is the reduced consumption of memory on the GPU devices. All of the storage arrays required by the simulation must be declared within the device memory on the GPU. The 16GB memory capacity of the Nvidia P100 GPUs is roughly an order of magnitude smaller than that found on a modern CPU node. This is a significant restriction in the context of large-scale DNS. The \texttt{StoreSome} algorithm reduces the total number of storage arrays from 65 to 32 for the TGV problem. This effectively doubles the maximum allowed grid size for a given number of GPUs. 

\subsection{Parallel GPU scaling}
\begin{figure}
  \center
  \includegraphics[width=1.0\columnwidth]{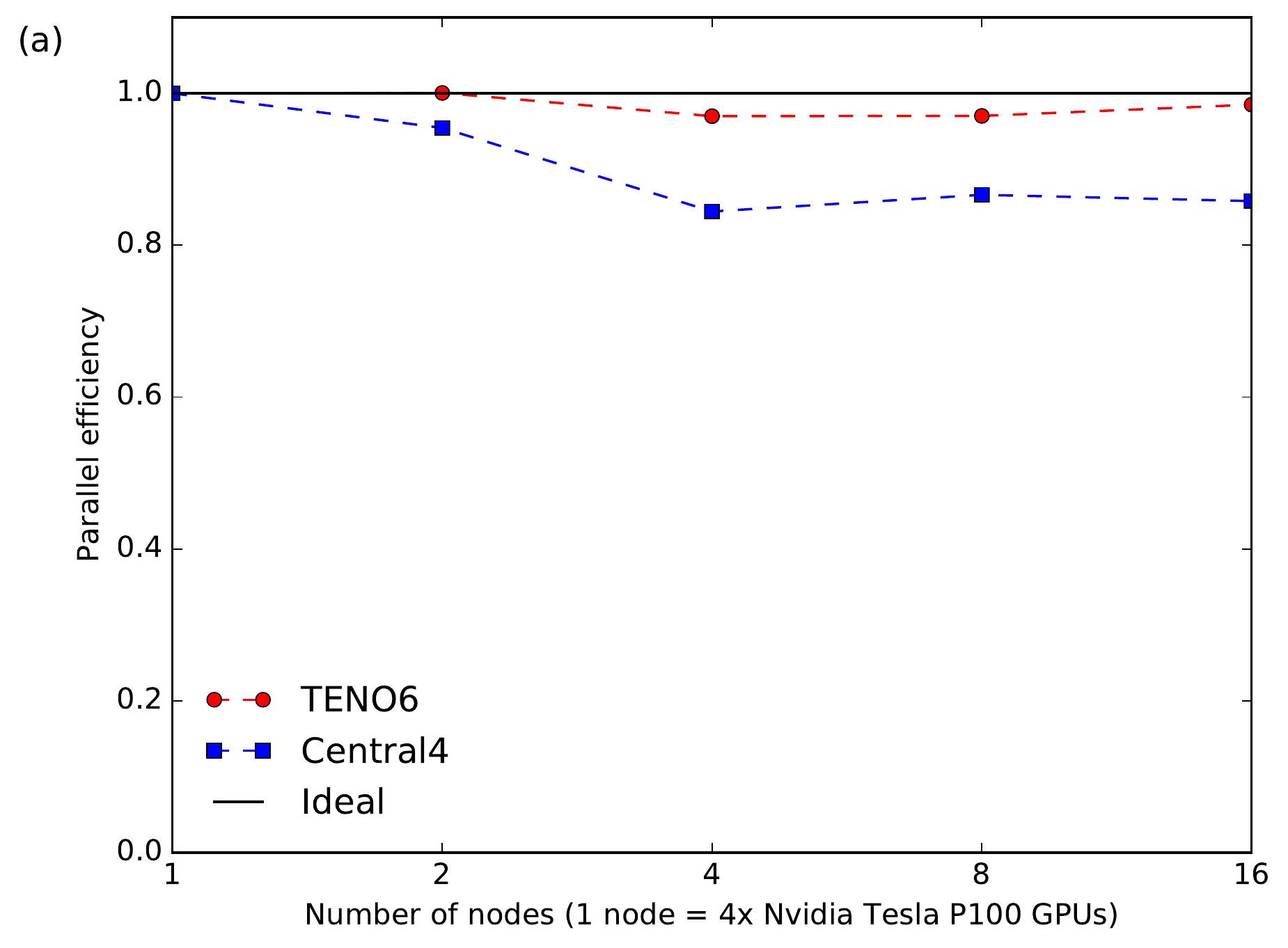}
  \includegraphics[width=1.0\columnwidth]{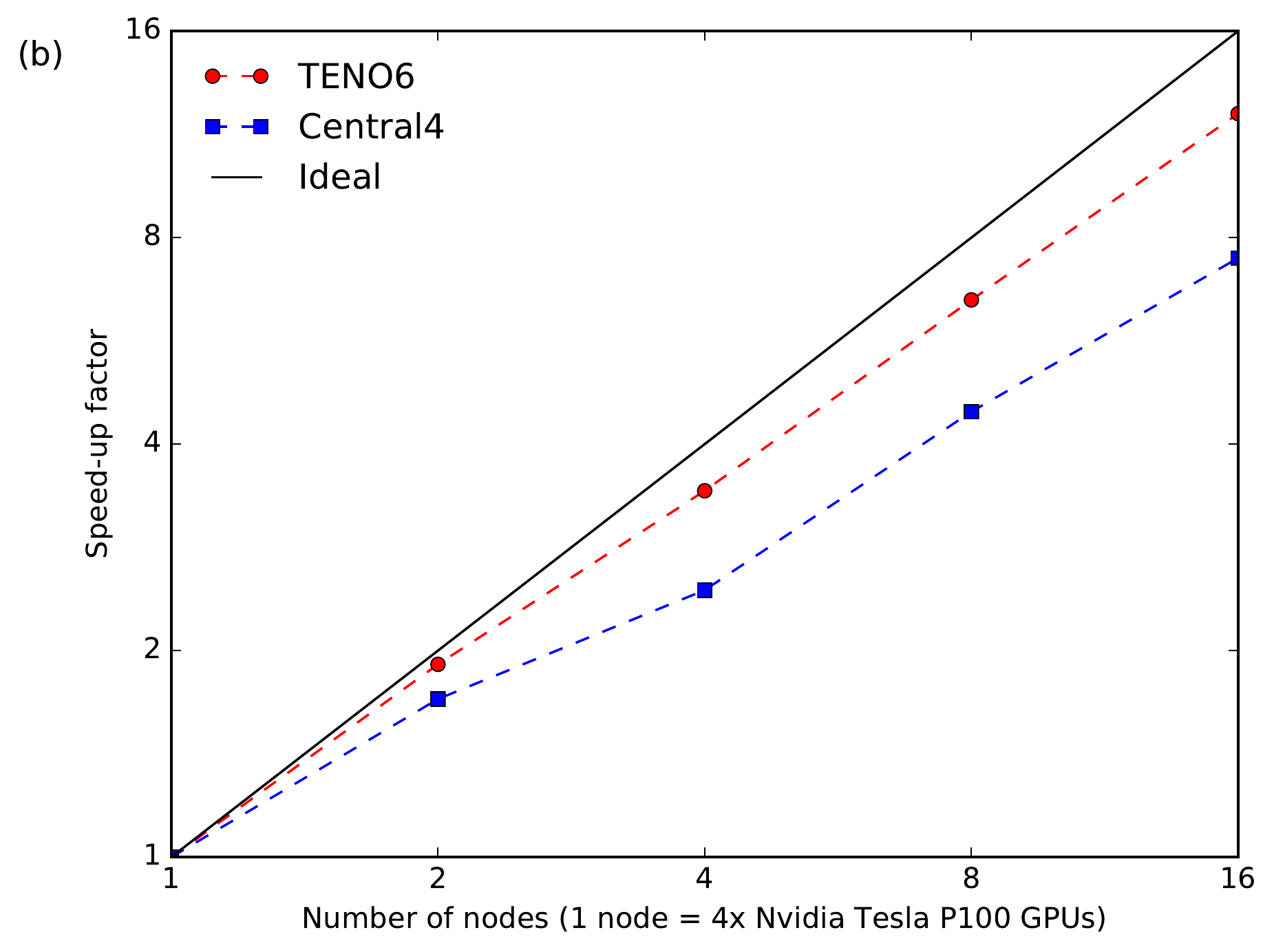}
  \caption{OpenSBLI scaling results for the \nth{4} order central and \nth{6} order TENO schemes on up to 64 GPUs with MPI+CUDA. (a) Weak scaling from 4 to 64 GPUs (b) Strong scaling from 4 to 64 GPUs for a $N=512^3$ grid.}\label{fig:scaling}
\end{figure}

Once the memory capacity of a single GPU is exceeded, the problem must be split across multiple devices. The host CPU nodes are then responsible for performing MPI halo exchanges between the decomposed memory blocks. This must be implemented efficiently to achieve good performance scaling over a large number of devices. In this section the weak and strong scaling of OpenSBLI/OPS is demonstrated for the TGV problem with the \nth{4} order central and \nth{6} order TENO schemes.

Figure \ref{fig:scaling} (a) shows the weak scaling of OpenSBLI on 4 to 64 GPUs. The base grid of $N=512^3$ points on 4 GPUs is doubled with each doubling in GPU count. The central scheme achieves good weak scaling, with around 85\% parallel efficiency at the largest grid size. There is a dip at the 16 GPU point, likely caused by the $(4\times 2\times 2)$ MPI decomposition proving suboptimal for the cubic grid distribution. The TENO6 scheme shows excellent weak scaling of 95\%, due to the increased operation counts of shock-capturing schemes. Figure \ref{fig:scaling} (b) shows the strong scaling of OpenSBLI on 4 to 64 GPUs. The runtime improvement is measured for increasing system resources, from an initial grid of $N=512^3$ on 4 GPUs. The central scheme achieves a 50\% speed-up on the largest GPU configuration, and shows a similar dip in scaling at 16 GPUs as observed in the weak scaling test. The \nth{6} order TENO scheme scales well, reaching 75\% of the perfect linear strong scaling.

\section{Conclusions}\label{sec:conclusions}
This work has described the OpenSBLI code-generation system for compressible fluid dynamics on heterogeneous computing architectures. Based on an earlier proof of concept \cite{JACOBS201712}, the new code incorporates high-order shock-capturing schemes, curvilinear coordinate transformations, and a wide range of boundary conditions. The Python-based code-generation system generates code in the OPS domain specific language \cite{Reguly_2014_OPSC}, enabling parallel execution on a wide range of computational hardware. The design and main components of the system have been discussed, with code examples demonstrating how the code is used in practice.

A suite of validation and verification cases has been presented, selected to demonstrate specific parts of the solver. A 3D DNS of a transitional shockwave/boundary-layer interaction was used to highlight the feasibility of code-generation for complex fluid flow problems. OpenSBLI was shown to exhibit good weak and strong scaling on multiple GPUs, highlighting its suitability for large-scale DNS. Additionally, a low-memory intensity algorithm in the code demonstrated the performance benefit of reduced global memory access. Future work will extend the capability of the code to multi-block problems, and to applications beyond the compressible Navier-Stokes equations.

\subsection*{Acknowledgements}
The authors would like to acknowledge the contributions of Dr. Gihan Mudalige and Dr. Istvan Reguly to this project, for the development and continued support of the Oxford Parallel Structured (OPS) library used in this work. The authors would also like to thank the following people for their testing and valuable feedback on the OpenSBLI code: Julien Lefieux (ONERA), Alex Gillespie and Hiten Mulchandani (University of Southampton), Dr. Arash Hamzehloo (Imperial College London), Dr. Gary Coleman (NASA Langley), and Dr. Andrea Sansica (JAXA).

David J. Lusher was funded by an EPSRC Centre for Doctoral Training grant (EP/L015382/1). Compute resources used in this work were provided by the `Cambridge Service for Data Driven Discovery' (CSD3) system operated by the University of Cambridge Research Computing Service (\url{http://www.hpc.cam.ac.uk}) funded by EPSRC Tier-2 capital grant EP/P020259/1, and the IRIDIS5 High Performance Computing Facility, and associated support services at the University of Southampton. The OpenSBLI code is available at \url{https://github.com/opensbli}. Data from this report will be available from the University of Southampton institutional repository. Declaration of Interests. The authors report no conflict of interest.





\bibliographystyle{elsarticle-num}
\bibliography{V2_paper.bib}







\end{document}